\documentclass[aps,english,floatfix,
tightenlines,twocolumn,nofootinbib,superscriptaddress]{revtex4-2}

\usepackage{amssymb}
\usepackage{amsmath,bm}
\usepackage{amssymb}
\usepackage{graphicx}
\usepackage{amsfonts}         
\usepackage{fancybox}

\usepackage[usenames,dvipsnames]{xcolor}%http://en.wikibooks.org/wiki/LaTeX/Colors

\usepackage[pagebackref,  % this puts links to the page numbers where refs appear
bookmarks={false}, pdfauthor={John McGreevy}, pdftitle={Yay, physics!}]{hyperref}
\hypersetup{colorlinks=true, 
linkcolor=BrickRed, 
citecolor=Violet, 
filecolor=OliveGreen, 
urlcolor=RoyalBlue, 
filebordercolor={.8 .8 1}, 
urlbordercolor={.8 .8 0}}%http://en.wikibooks.org/wiki/LaTeX/Hyperlinks
\usepackage{soul}%strike through text \st{} .  not in math mode.
\setstcolor{Red}
\definecolor{darkgreen}{rgb}{0,0.4,0}
\definecolor{darkred}{rgb}{0.4,0,0}
\definecolor{darkblue}{rgb}{0,0,0.4}
\definecolor{lightblue}{rgb}{.6,.6,0.9}

\newcommand{\cobl}{\color{blue}}

\definecolor{uglybrown}{rgb}{0.8,  0.7,  0.5}

\definecolor{palatinatepurple}{rgb}{0.41, 0.16, 0.38}
\definecolor{celebrationcolor}{rgb}{0.75,  0.0,  0.9}

\definecolor{Mathematica1}{rgb}{0.368417, 0.506779, 0.709798}
\definecolor{Mathematica2}{rgb}{0.880722, 0.611041, 0.142051}
\definecolor{Mathematica3}{rgb}{0.560181, 0.691569, 0.194885}

\definecolor{shadecolor}{rgb}{0.90,0.90,0.90}
\definecolor{DVcolor}{rgb}{0.95,  0.5,  0.2}
\definecolor{lightbluemuons}{rgb}{0.0,.65,1.0}

\definecolor{chartreuse}{rgb}{0.70, 1.00, 0.00}

\usepackage{tikz}
\usetikzlibrary{shapes}
\usetikzlibrary{trees}
\usetikzlibrary{snakes}
\usetikzlibrary{matrix,arrows} 				% For commutative diagram
\usetikzlibrary{positioning}				% For "above of=" commands
\usetikzlibrary{calc,through}				% For coordinates
\usetikzlibrary{decorations.pathreplacing}  % For curly braces
\usepackage[tikz]{bclogo} 					% For cute logo boxes
\usepackage{pgffor}							% For repeating patterns

\usetikzlibrary{decorations.markings}
\usetikzlibrary{intersections}				% For interesections

\usetikzlibrary{arrows,decorations.pathmorphing,backgrounds,positioning,fit,petri,automata,shadows,calendar,mindmap, graphs}
\usetikzlibrary{arrows.meta,bending}

\tikzset{
    vector/.style={decorate, decoration={snake}, draw},
    fermion/.style={postaction={decorate},
        decoration={markings,mark=at position .55 with {\arrow{>}}}},
    fermionbar/.style={draw, postaction={decorate},
        decoration={markings,mark=at position .55 with {\arrow{<}}}},
    fermionnoarrow/.style={},
    gluon/.style={decorate,
        decoration={coil,amplitude=4pt, segment length=5pt}},
    scalar/.style={dashed, postaction={decorate},
        decoration={markings,mark=at position .55 with {\arrow{>}}}},
    scalarbar/.style={dashed, postaction={decorate},
        decoration={markings,mark=at position .55 with {\arrow{<}}}},
    scalarnoarrow/.style={dashed,draw},
	vectorscalar/.style={loosely dotted,draw=black, postaction={decorate}},
}

\def\centerarc[#1](#2)(#3:#4:#5)% Syntax: [draw options] (center) (initial angle:final angle:radius)
    { \draw[#1] ($(#2)+({#5*cos(#3)},{#5*sin(#3)})$) arc (#3:#4:#5); }

\usepackage{enumitem}

\usepackage{slashed}

\usepackage[font=small,labelfont=bf]{caption}
\DeclareCaptionFont{tiny}{\tiny}
\usepackage{epstopdf}
\usepackage{wasysym}

\usepackage[yyyymmdd,hhmmss]{datetime}   
\usepackage{framed}  % needed for \begin{shaded}
 \usepackage{mdframed}
 \usepackage{wrapfig}
 \usepackage{yfonts}  
\newmdenv[%
        backgroundcolor=lightgray,
    linecolor=black,
    outerlinewidth=2pt,
]{boxedandshaded}

\def\parfig#1#2{
\parbox{#1\textwidth}
{\includegraphics[width=#1\textwidth]{#2}}
}

\numberwithin{equation}{section}

\renewcommand{\theequation}{\arabic{section}.\arabic{equation}}

\newcommand{\vev}[1]{\langle #1 \rangle}

\newlength{\extraspace}
\newlength{\extraspaces}
\def\be{\begin{equation}}
\def\ee{\end{equation}}

\newcommand{\bea}{\begin{eqnarray}}
\newcommand{\eea}{\end{eqnarray}}

\def\half{{1\over 2}}

\def\bra#1{\left\langle#1\right|}
\def\ket#1{\left|#1\right\rangle}

\def\vev#1{\left\langle{#1}\right\rangle}

\def\II{\relax{I\kern-.10em I}}

\def\IB{\relax{\rm I\kern-.18em B}}

\def\ID{\relax{\rm I\kern-.18em D}}
\def\IE{\relax{\rm I\kern-.18em E}}
\def\IF{\relax{\rm I\kern-.18em F}}
\def\IG{\relax\hbox{$\inbar\kern-.3em{\rm G}$}}
\def\IGa{\relax\hbox{${\rm I}\kern-.18em\Gamma$}}
\def\IH{\relax{\rm I\kern-.18em H}}
\def\II{\relax{\rm I\kern-.18em I}}
\def\IK{\relax{\rm I\kern-.18em K}}
\def\inbar{\,\vrule height1.5ex width.4pt depth0pt}

\def\simgt{\hskip0.05in\relax{ 
\raise3.0pt\hbox{ $>$
{\lower5.0pt\hbox{\kern-1.05em $\sim$}} }} \hskip0.05in}

\def\lp10{\ell_p^{10}}
\def\lp11{\ell_p^{11}}
\def\R11{R_{11}}

\def\frac#1#2{{#1 \over #2}}

\newdimen\tableauside\tableauside=1.0ex
\newdimen\tableaurule\tableaurule=0.4pt
\newdimen\tableaustep
\def\phantomhrule#1{\hbox{\vbox to0pt{\hrule height\tableaurule width#1\vss}}}
\def\phantomvrule#1{\vbox{\hbox to0pt{\vrule width\tableaurule height#1\hss}}}
\def\sqr{\vbox{%
  \phantomhrule\tableaustep
  \hbox{\phantomvrule\tableaustep\kern\tableaustep\phantomvrule\tableaustep}%
  \hbox{\vbox{\phantomhrule\tableauside}\kern-\tableaurule}}}
\def\squares#1{\hbox{\count0=#1\noindent\loop\sqr
  \advance\count0 by-1 \ifnum\count0>0\repeat}}
\def\tableau#1{\vcenter{\offinterlineskip
  \tableaustep=\tableauside\advance\tableaustep by-\tableaurule
  \kern\normallineskip\hbox
    {\kern\normallineskip\vbox
      {\gettableau#1 0 }%
     \kern\normallineskip\kern\tableaurule}%
  \kern\normallineskip\kern\tableaurule}}
\def\gettableau#1 {\ifnum#1=0\let\next=\null\else
  \squares{#1}\let\next=\gettableau\fi\next}

\def\({\left(}
\def\){\right)}

\def\lsim{\mathrel{\mathstrut\smash{\ooalign{\raise2.5pt\hbox{$<$}\cr\lower2.5pt\hbox{$\sim$}}}}}
\def\gsim{\mathrel{\mathstrut\smash{\ooalign{\raise2.5pt\hbox{$>$}\cr\lower2.5pt\hbox{$\sim$}}}}}

\def\overleftrightarrow#1{\vbox{\ialign{##\crcr
     $\leftrightarrow$\crcr\noalign{\kern-0pt\nointerlineskip}
     $\hfil\displaystyle{#1}\hfil$\crcr}}}
     
     \def\overleftarrow#1{\vbox{\ialign{##\crcr
     $\leftarrow$\crcr\noalign{\kern-0pt\nointerlineskip}
     $\hfil\displaystyle{#1}\hfil$\crcr}}}

\def\eg{{\it e.g.}}

\newif{\ifeq}           % defines a new condition @eq tested by the conditional \ifeq
\eqtrue                 % if uncommented, declares @eq to be true
\newcounter{lecturecounter}
\usepackage{array} %% for \extrarowheight
\usepackage{comment}
\usepackage{cleveref}
\usepackage{makecell}

\usepackage[makeroom]{cancel}

\usepackage{lipsum}
\renewcommand{\theequation}{\arabic{equation}}

\numberwithin{equation}{section}
\renewcommand{\theequation}{\arabic{section}.\arabic{equation}}

\definecolor{XLgreen}{RGB}{34,139,34}
\definecolor{JMblue}{RGB}{25,25,125}
\definecolor{BSorange}{RGB}{140,50,0}

\def\high{\Delta}
\newcommand{\<}{\langle}
\renewcommand{\>}{\rangle}
\def\cd{c_{\Delta}}
\def\hcd{\hat{c}_{\Delta}}

\begin{document}

%\title{A systematic search for conformal field theories in very small systems}

\title{A systematic search for conformal field theories in very small spaces}

%\title{A systematic search for conformal field theories in very small Hilbert spaces}

%
\author{Xiang Li}
\affiliation{Department of Physics, University of California at San Diego, La Jolla, CA 92093, USA}
\author{Ting-Chun Lin}
\affiliation{Department of Physics, University of California at San Diego, La Jolla, CA 92093, USA}
\affiliation{Hon Hai Research Institute, Taipei, Taiwan}
\author{John McGreevy}
\affiliation{Department of Physics, University of California at San Diego, La Jolla, CA 92093, USA}

\begin{abstract}
Groundstates of 1+1d conformal field theories (CFTs) satisfy a local entropic condition called the vector fixed point equation.  This condition is surprisingly well satisfied by groundstates of quantum critical lattice models even at small system sizes.  We perform a search in the space of states of very small systems (four qubits and four qutrits) and examine the states that satisfy this condition.  By reconstructing a local Hamiltonian from each state, we are able to identify many of these solutions with known CFTs; others are gapped fixed points, or involve large relevant perturbations, and others are CFTs we have not yet identified.  These ideas are also useful for identifying continuous quantum phase transitions in a given family of Hamiltonians, and for identifying the nature of the critical theory in small systems.

\end{abstract}

%\today
\date{\today}

\maketitle

\section{Introduction}

Conformal field theories (CFTs) are the endpoints of renormalization group flows of relativistic quantum field theories, govern the universal information at continuous phase transitions, and in the case of 1+1 dimensions, are a crucial ingredient in perturbative string vacua.  The space of CFTs remains an object of mystery.  Even in 1+1 dimensions, where conformal symmetry implies that a theory represents an infinite-dimensional Virasoro algebra \cite{Belavin:1984vu}, a classification of unitary CFTs only exists when the Virasoro central charge $c$ is less than 1 \cite{Friedan:1983xq}.  Our knowledge of CFTs with $c \geq 1$ relies on ad hoc constructions, and solvable models with extra symmetries (see e.g.~\cite{DiFrancesco:1997nk}).  As a result, all well-understood explicit constructions produce CFTs that are rational with respect to some chiral algebra (Virasoro or one of its extensions).  Despite a general belief that the generic CFT is irrational, and evidence from holographic duality for the existence of many irrational CFTs, an explicit example of an irrational CFT is not yet known with certainty (but see \eg~\cite{Dotsenko:1998gyp, Dotsenko:2001cct, Antunes:2022vtb, Antunes:2024mfb,Antunes:2025huk} and references therein for a candidate construction of a few such theories).

The Entanglement Bootstrap is a program to understand the universal properties of quantum many-body states from their entanglement structure, and it provides a new perspective on quantum field theory (QFT).
Let us define a universal property of a state to be a property of a renormalization group fixed point, shared by its whole basin of attraction.
%The idea is that the universal information in a groundstate is encoded in the low-energy QFT. In the case of quantum critical lattice models, this low-energy QFT is a CFT.
The idea is that the universal information in a groundstate is that of its low-energy QFT. In the case of quantum critical lattice models, this low-energy QFT is a CFT.

The thus-far-most-well-developed aspect of Entanglement Bootstrap is its application to liquid gapped bosonic phases
\cite{Shi:2018krj, Shi:2018bfb,Shi:2019mlt, Shi:2019ngw,Shi:2020jxd, Shi:2020rne,knots-paper,paper-one}, where the low-energy field theory is a topological quantum field theory (TQFT).
%Much more mysterious are gapless states, and in particular conformal field theories (CFTs).
%Even in 1+1 dimensions, there is no classifcation of CFTs, and while the generic CFT is believed to be irrational, no explicit example is known.
A key ingredient of Entanglement Bootstrap for gapped states is a set of local conditions on the state that guarantee that the state is a good representative of a gapped liquid topological order.  From these local conditions, much of the machinery of TQFT has been constructed.

In \cite{Lin:2023pvl}, we proposed an analogous local condition for 1+1 dimensional CFT groundstates,
\begin{align}
\label{eq:VFPE}
K_\Delta \ket{\psi}  \propto \ket{\psi},
&~~
K_\Delta \equiv \eta \hat \Delta(A,B,C)
+ (1-\eta) \hat I(A:C|B),
\\
~\hat \Delta(A,B,C) &\equiv
K_{AB} + K_{BC} - K_A - K_C,
\\ ~\hat I(A:C|B)  & \equiv K_{AB} + K_{BC} - K_B  - K_{ABC}
\end{align}
where $A,B,C$ are three adjacent intervals, and $\eta$ is the cross-ratio specified by their four endpoints.   The fact that a CFT groundstate satisfies this condition,
which we call the vector fixed-point equation (VFPE), follows from known properties of the CFT groundstate entanglement Hamiltonian of an interval \cite{Casini:2011kv, Cardy:2016fqc}.
Conversely, there is strong evidence that a state satisfying the VFPE actually must be a fixed point of the renormalization group\footnote{We note that not all such fixed points are CFTs; some of them have zero correlation length, rather than infinite correlation length.  We will explain below how to distinguish these two classes of states.}.
Amongst the evidence for this conjecture, we mention
its strong consequences for the form of the entanglement Hamiltonian \cite{strength-of-vfpe},
that a reformulation of this condition \cite{Kim:2024suq} can logically result in the emergence of conformal geometry,
as well as the ability to extract expressions for all six global conformal generators (including the Hamiltonian) and Virasoro generators \cite{Kim:2024upb}. 
This VFPE, although is derived from continuum CFT, works surprisingly well for critical lattice models, even in very small system sizes with small local Hilbert spaces. This is not true for all results derived from field theory, due to strong finite size effect.
%It is designed not just as an identity with entanglement Hamiltonians, but also in a way
The VFPE is designed in such a way that finite-size effects and UV physics are suppressed to some extent. This is the key reason that we can impose this condition on small systems to study universal properties of phases of matter.

The VFPE turns out \cite{Lin:2023pvl} to be the condition that the state is a critical point of a certain linear combination of entropies of three contiguous regions, $S_\Delta \equiv \vev{\psi| K_\Delta | \psi } $.  This suggests the possibility that $S_\Delta$ can behave like a Morse function on the space of 1+1d states, similarly to the famous Zamolodchikov $c$-function \cite{Zamolodchikov}, which is instead defined (for relativstic QFT) in terms of correlation functions of local operators, and in terms of which the renormalization group is gradient flow. The relationship between our entropy function and the renormalization group is still open, but this picture has an exciting practical consequence: it suggests a systematic algorithm to enumerate CFTs.
A nice feature of this algorithm is that it requires no input about symmetry or of a choice of Hamiltonian. In particular, there is no reason that CFTs found this way should be rational for some chiral algebra.

In this paper, we take steps toward implementing this algorithm.  A crucial ingredient is the ability \cite{Lin:2023pvl} to reconstruct a local Hamiltonian from the groundstate.
The reconstructed Hamiltonian, on a circle with $L$ sites, is
\be\label{eq:Hrec} H_\text{rec}^\psi
= {\cot { \pi \over L } \over L} \sum_i \( K_{i,i+1} - K_{i}\)  . \ee
where $K_X$ is the entanglement Hamiltonian for region $X$ from the state $\ket{\psi}$.
This Hamiltonian has several virtues.
First, it is a local Hamiltonian; in fact it involves only nearest-neighbor interactions.
This means that from a small system, we can reconstruct the Hamiltonian for a large system.
Second, $H_{\text{rec}}$ is a non-negative operator. This can be seen by write it as
\be H_\text{rec}^\psi = {\cot { \pi \over L } \over 2 L}\sum_i \hat \Delta(i-1, i, i+1),
\ee
%(where $\hat \Delta(x_1,x_2,x_3,x_4) \equiv \hat \Delta(A,B,C)$ with the endpoints of $A,B,C$ specified by the arguments) \XL{We can just treat $i$ as a label of interval, then it is just $\hat{\Delta}(i-1,i,i+1)$. This is consistent with the label $K_{i,i+1}, K_i$ where $i$ is the interval.}
% because each 
where $\hat \Delta \geq 0$ by operator weak monotonicity \cite{Lin:2022jtx}.
Third, if the state is the groundstate of a CFT, and each site is identified with an interval in the continuum, it is exactly the Hamiltonian of the CFT on the circle, with the normalization chosen so that the eigenvalues are $2\pi \Delta_i/L$
where $\{\Delta_i\}$ are the CFT scaling dimensions,
\begin{equation}
\label{eq:Hrec-and-dimensions}
  (H_\text{rec}^\psi - E_0) |\Delta_i\> = \frac{2\pi}{L} \Delta_i |\Delta_i\>,
\end{equation}
where $E_0$ is the ground state energy.
Elsewhere \cite{strength-of-vfpe} we will show that
a state $\ket{\psi}$ satisfying the fixed point equation
%\eqref{eq:VFPE}
is the groundstate of its reconstructed Hamiltonian
$H_\text{rec}^\psi$\footnote{
  The assumptions required to show the groundstate result are stronger than those used to enumerate CFT in this paper.
  Specifically, proving the groundstate result requires taking the limit of many sites,
    $L \to \infty$,
  and demands that the fixed-point equation hold for all possible three intervals.
  Whereas in this paper, we rely only on the case $L = 4$.
}.
The spectrum of $H_\text{rec}$ computed from the groundstates of a critical quantum lattice model matches surprisingly well with that of the original Hamiltonian even for small system sizes.
The fixed point equation for $\ket{\psi}$ also implies $z=1$ scale invariance of $H_\text{rec}^\psi$ \cite{Lin:2023pvl}.

\section{Strategy for enumeration of CFTs}
\def\>{\rangle}
\def\<{\langle}

\begin{figure}
  \centering
  \includegraphics[clip,trim=7cm 4cm 7cm 5cm,width=\linewidth]{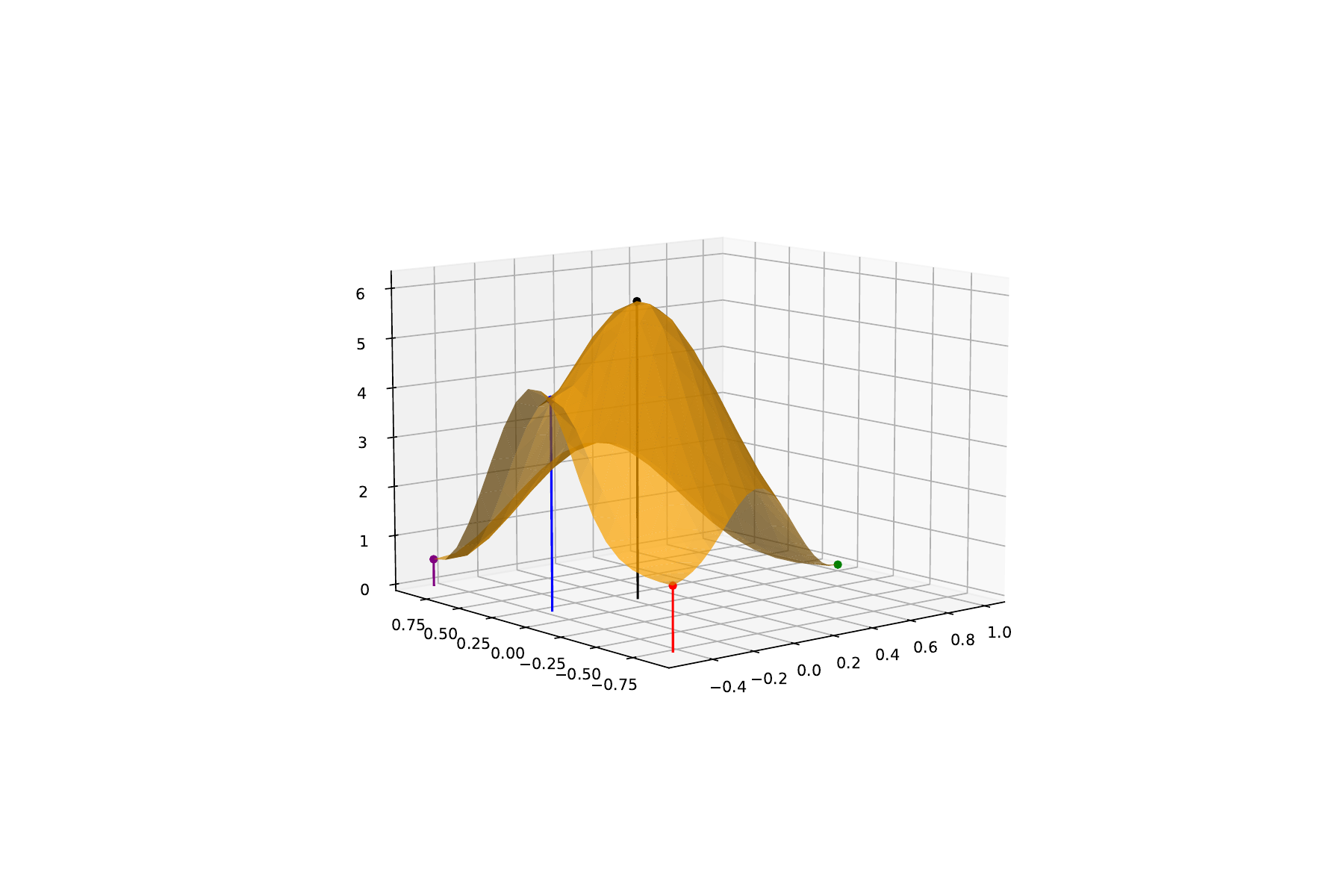}
  \caption{A two-dimensional cross section of the function $S_\Delta$ on the space of translation-invariant real wavefunctions on four qubits.  Remarkably, five of the critical points listed in Table \ref{table:critical-4-qubit-states} can be seen, including
  trivial (green),
  Ising (purple),
  Heisenberg (red),
  Fibonacci (blue),
  and maximum (black).}
  \label{fig:morse-function}
\end{figure}

With confidence from the numerical tests in \cite{Lin:2023pvl, Ozzello:2025tfu}, we are ready to describe an algorithm that potentially enumerates all unitary CFTs.
The algorithm contains two parts.
The first part generates approximate CFT ground states $|\psi\>_{ABCD}$ on a circle of $4$ evenly partitioned intervals, $A, B, C, D$.
We replace each interval with a single `site', a Hilbert space of finite dimension, $d$, which play the role of a control parameter in our search.
We proceed by finding states that are the critical points of $S_\Delta$ with $\eta = 1/2$.
We search within the set of translation-invariant states on four sites.
%Note that in this situation, $K_\Delta \ket{\psi} \propto  H_\text{rec} \ket{\psi}$ using purity of the state (\eg~$K_{ABC} \ket{\psi} = K_D \ket{\psi}$).
For now, we also restrict our search to time-reversal-invariant, i.e.~real, states (see \Cref{fig:morse-function}).

The second part of the procedure generates approximate CFT Hamiltonians from the ground states using \eqref{eq:Hrec}.
Once we have the Hamiltonian, we can obtain the CFT data, including scaling dimensions and OPEs, through known algorithms \cite{zou2020conformal}.
Note that the purpose of this algorithm is not to give accurate values for the CFT data.
The purpose is to generate as many CFTs as possible, including previously unknown ones.
If one wants accurate values, improvements to the Hamiltonians can be made,
%improvements to the algorithm can be made,
and conformal bootstrap \cite{poland2019conformal} can be used afterward to turn these approximate values into tighter bounds.

Before continuing,
let us introduce the operator $\hat{c}_\Delta$ proportional to $H_\text{rec}$:
\begin{multline}
  \hat{c}_\Delta(\psi) \equiv \frac{3}{2 \log 2} (K_{AB} - K_B + K_{BC} - K_C \\
  + K_{CD} - K_D + K_{DA} - K_A)
\end{multline}
with expectation value
\begin{equation}
  c_\Delta(\psi) \equiv \<\psi| \hat{c}_\Delta(\psi) |\psi\>.
\end{equation}
Since $\hat{c}_\Delta \,\propto\, H_\text{rec}$,
we have $\hat{c}_\Delta |\psi\> \,\propto\, H_\text{rec} |\psi\>  \,\propto\, |\psi\>$ for CFT groundstates.
The advantage of this normalization is that when $|\psi\>$ is the ground state of the field theory,
  $c_\Delta$ is exactly the central charge of the corresponding CFT.
Thus, equivalently, the goal in the first part is to find all the critical points of $c_\Delta$.

{\bf Search.}
The process of finding the critical point consists of two steps.
First, we find an initial point close to the critical point.
 Notice that we cannot just do gradient descent on $c_\Delta$, because we want to find all critical points, not just the local minima.\footnote{If instead the goal is to find CFTs with spectral gap $> 2$ (also known as dead-end CFTs, since they cannot be perturbed by relevant operators),
then it may be sufficient to study the local minima.}
Generally, performing gradient descent only gives local minima. On the other hand, it is indeed possible to perform gradient descent over the error function\footnote{Note that when acting on a translation-invariant state, $H_\text{rec}$ and $K_\Delta$ are proportional to each other by a purification move (\eg~$K_{ABC} \ket{\psi} = K_D \ket{\psi}$).}, which is the standard deviation of $\hat{c}_\Delta$\footnote{
 Note that our choice of normalization differs from that used in \cite{Lin:2023pvl}.
 }:
\be \text{err} \equiv \sigma(\hat{c}_\Delta) \equiv
 \sqrt{ \bra{\psi} \( \hat{c}_\Delta(\psi) \)^2 \ket{\psi}
 - \bra{\psi} \hat{c}_\Delta(\psi) \ket{\psi}^2} . \ee
Second, we use a second-order Newton's method to converge to the critical point.
Some further details of the search procedure are given in App.~\ref{app:search-details}.

% \david{i moved some of this discussion in the analysis}
%\JM{I don't see where this appears and I think it is important to say: }
Note that $c_\Delta$ is invariant under local unitary transformations.  Thus, each critical point is actually an orbit of this gauge redundancy.
% Because we focus on real states this redundancy is less severe;
% we fix it up to some discrete freedom below.
%it is only a discrete redundancy, which we fix below.
%\david{it is not a discrete redundancy right? because we have $O(2)$ or $O(3)$.}
%\david{also we do not completely fix}

{\bf Analysis.} Once the critical points are identified, we perform a series of diagnostics to determine whether their properties are consistent with those of a conformal field theory.

As a preliminary step, we discard states with degenerate ground states,
  since it is believed that a CFT has a unique ground state.

Next, we check whether the low-energy spectrum of the reconstructed Hamiltonian $H_\text{rec}$ exhibits universal behavior as the system size $L$ increases.
Let $E_{L, i}$ denote the $i$-th eigenvalue of $H_\text{rec}$ on $L$ sites.
In the absence of relevant perturbations, one expects that for fixed $i$, the limit $\lim_{L \to \infty} L \cdot E_{L, i}$ exists. In fact this condition is also satisfied even if a small relevant perturbation is present, as illustrated in In Fig.~\ref{fig:effects-of-relevant-perturbation}. 
Thus, the diagnostic is to test whether the low-lying spectrum appears to decrease at the rate $1/L$ for $L$ that does not exceed the correlation length.
%Thus, the diagnostic is to test whether the spectrum appears to converge as $L$ grows.
%In Fig.~\ref{fig:effects-of-relevant-perturbation} we illustrate the effect of a relevant perturbation on the spectrum.

We also examine whether the ground state exhibits the entanglement entropy scaling expected of a CFT, namely
$S(\ell) = \frac{c}{3} \log \sin \frac{\pi \ell}{L} + \text{const}$.
The diagnostic is to test whether a plot of $S(\ell)$ vs $\log \sin \frac{\pi \ell}{L}$ yields a straight line.

In both cases, we use periodic MPS to obtain the low-energy spectrum and corresponding eigenstates.

\section{Results}

\begin{figure}
\parfig{.5}{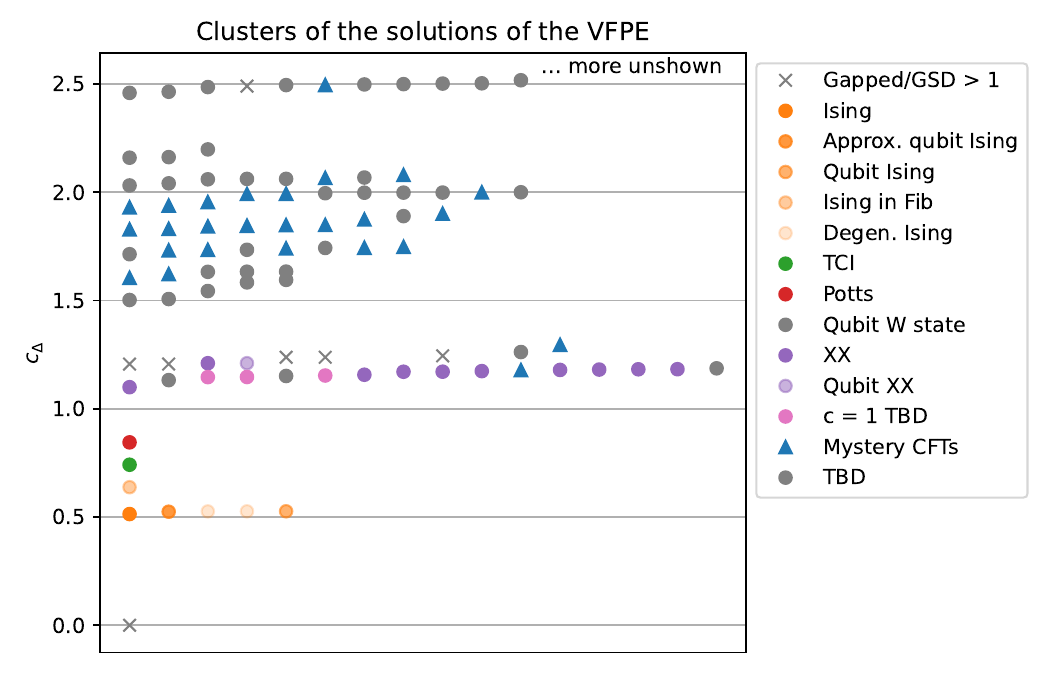}
  \caption{Critical 4-qutrit states.
  We do not display the states with larger values of $c_\high$.  ``Degen.~Ising" refers to a state with the spectrum of the critical Ising model, but with two-fold degeneracy.  ``Ising in Fib." refers to a state whose spectrum on $L$ sites is the first $\ell_L$ states of the Ising model below a gap, where $\ell_L$ is the $L$th Lucas number.
  States labelled `gap' are gapped fixed points.
$H_\text{rec}$ of the states marked with an $x$
have a groundstate degeneracy on 4 sites and were discarded.
  }
  \label{fig:critical-4-qutrit-states}
\end{figure}

\Cref{table:critical-4-qubit-states} and \Cref{fig:critical-4-qutrit-states} list the critical states we found over 4 qubits and 4 qutrits respectively.
%\david{fill when we get a more complete list of qutrit states}
In the case of qutrits, we perform sampling --  it might not be a complete list, but a list of solutions we have found.
In both cases, we restrict our search to translation-invariant and time-reversal-invariant states.
For qubits and qutrits, respectively, this space has 5 and 23 dimensions. 

\begin{table}
\begin{center}
  \begin{tabular}{|c|c|c|}
  \hline
  $c_\high$ & Description & Explicit form up to on-site unitary\\
  \hline
  \hline
  0 & cat states & $a|0000\> + b|1111\>$ \\
  \cobl  0.526 &   \cobl Ising CFT &
  $a_1|0000\> + a_2|1111\>
  + a_5\ket{\alpha} + a_6 \ket{\beta} $
  \\
  1.132 & W state & $\frac{1}{2}(|0001\> + |0010\> + |0100\> + |1000\>)$ \\
  \cobl  1.211 &   \cobl XX model & \makecell{$-\frac{1}{\sqrt{2}} \ket{\beta} + \frac{1}{\sqrt{2}} \ket{\alpha}$} \\
  \cobl  1.245 &   \cobl Heisenberg & \makecell{$-\sqrt{\frac{2}{3}} \ket{\beta}  + \frac{1}{\sqrt{3}} \ket{\alpha}$} \\
  3.510 & ferromagnet & \makecell{$\frac{1}{\sqrt{3}}|0000\> +$ \\ $\frac{1}{\sqrt{6}}(|0111\> + |1011\> + |1101\> + |1110\>)$}\\
  4.165 & Fibonacci chain & $\frac{1}{\sqrt[4]{5}}\Big(|0000\> + \sqrt{\sqrt{5}-1}\ket{\beta}\Big)$\\
  6.000 & maximum point & $\frac{1}{2}(|0000\> \pm |1111\> + |0101\> + |1010\>)$ \\ % full degeneracy
  \hline
  \end{tabular}
\end{center}
  \caption{
  All critical 4-qubit states. Actual CFTs are in blue.  We defined
  $\ket{\alpha} \equiv ( \ket{0011} + \ket{0110} + \ket{1100} + \ket{1001} )/2$ and $\ket{\beta} \equiv ( \ket{1010} + \ket{0101})/\sqrt{2}$ as elements of our basis of translation-invariant states. The coefficients for the Ising solutions are
  $[a_1, a_2, a_5, a_6]
  = [0.11036, 0.80152, 0, 0, -0.51103, 0.29023].$
  In fact, there are two inequivalent maximum point state with the same $H_\text{rec}$ and $S_\Delta$ that are not related by on-site unitaries.
  }
  \label{table:critical-4-qubit-states}
\end{table}

{\bf Qubits.}
The entries in the table are all RG fixed points, but some of them are zero-correlation-length fixed points.
Actual CFTs are in blue\footnote{There exist translation-invariant states that are equivalent under an on-site unitary, but not under a translation-invariant on-site unitary.
For example, two realizations of the Ising CFT, $a_1|0000\> + a_2|1111\> + a_5\ket{\alpha} \pm a_6 \ket{\beta}$ are related by the unitary $I \otimes Z \otimes I \otimes Z$.}.
These two classes of states can be distinguished by reconstructing the local Hamiltonian from the 4-qubit state and studying it on larger system sizes. 
For the XX and Heisenberg solution, we discover that the reconstructed Hamiltonian are exactly the well-known Hamiltonians $H = XX + YY$ and $H = XX +YY +ZZ$ respectively up to a prefactor and an additive constant. In fact, this prefactor, which gives the exactly correct scaling behavior of the Hamiltonian, is a bonus from using the reconstructed Hamiltonian -- one does not need to manually scale the spectrum. 
The reconstructed Hamiltonians from the gapped fixed point states on qubits all have a groundstate degeneracy.

Some comments about the zero-correlation-length fixed points that appear in the table:
The groundstate degeneracy and $c_\high$ have a peculiar relation, i.e.
$\log \( \text{degeneracy}\) = 0.46 c_\high$.
There are only two eigenstates with finite energy for the reconstructed Hamiltonian of the W state: one ground state is the W state, and the other is the product state $|0\>^{\otimes n}$.
This is consistent with the theorem of
\cite{Gioia:2023adm}, which forbids a local Hamiltonian whose {\it unique} groundstate is the W state.
The reconstructed Hamiltonians for the states labelled `ferromagnet' and `Fibonacci'  are (up to an additive constant):
\begin{align}
  H_{\text{Ferromagnet}} &\propto -\sum_i \(X_i X_{i+1} + Y_i Y_{i+1} + Z_i Z_{i+1}\) \\
  H_{\text{Fibonacci}} & \propto \sum_i (|11\>\<11|)_{i i+1}
  \label{eq:H_fib}
\end{align}
which explains the nomenclature.
The latter is well-known to produce the golden chain
\cite{Feiguin:2006ydp, Aasen:2016dop,Aasen:2020jwb, Antunes:2025huk}, which
at system size $L=2,3,4,5,6,7,...$, has a ground state degeneracy $\ell_L= 3, 4, 7, 11, 18, 29,...$, where $\ell_L$ are the Lucas numbers.
The maximum point of $S_\Delta$ on four qubits is something like a perfect tensor,
 where $S_2 = 2 \log 2$ attains its maximal value for any two contiguous sites.

It is not a coincidence that the three CFTs that fit into the 4-qubit Hilbert space
in \Cref{table:critical-4-qubit-states} are ones for which we know a parent Hamiltonian acting on nearest-neighbor qubits.
Our reconstructed Hamiltonian always acts on 2 sites at a time.
The question of which CFTs will fit in a given local Hilbert space, with our nearest-neighbor reconstructed Hamiltonian, is closely related in spirit to the $c$-$d$ conjecture of
\cite{Latorre:2024uqi}, which says that a lattice model of local Hilbert space dimension $d$ and nearest-neighbor interactions can only realize a CFT with $c \leq d-1$.
The results so far are consistent with that conjecture.
In fact, our search strategy can in principle provide a rather stringent test of that conjecture:
In our search with local dimension $d$, we should find no CFTs with $c>d-1$.

{\bf Qutrits.}
Compared to
\Cref{table:critical-4-qubit-states},
when searching in the space of translation-invariant, real states on four qutrits, we find more CFTs (\Cref{fig:critical-4-qutrit-states}).
In the list for qutrits, there are multiple representatives of the Ising model.
One is the same state as realized on qubits, making no use of the third register (labelled `Ising qubits'), and the other is a new state whose value of $c_\high$ is closer to $\half$. Interestingly, there is a solution which encodes the Ising spectrum in the Fibonacci chain groundsate subspace.

We find representatives of the next two unitary minimal models after Ising, with
$c_\high = 0.744$ (tricritical Ising has $c = 7/10$) and
$c_\high = 0.844$ (3-state Potts  has $c=4/5$), as well as many more points along the $c=1$ line.
At larger $c$ we find many more (of order 25) states whose reconstructed spectrum looks like that of a CFT, possibly with small relevant perturbation.
In Appendix \ref{app:search-results-details} we give details of the information extracted from the solutions of the VFPE on qutrits.

\begin{figure}
\centering
\includegraphics[width=0.45\linewidth]{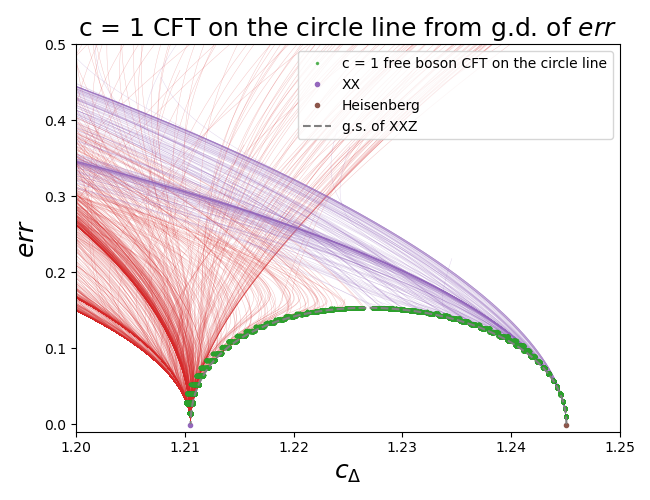}
\includegraphics[width=0.45\linewidth]{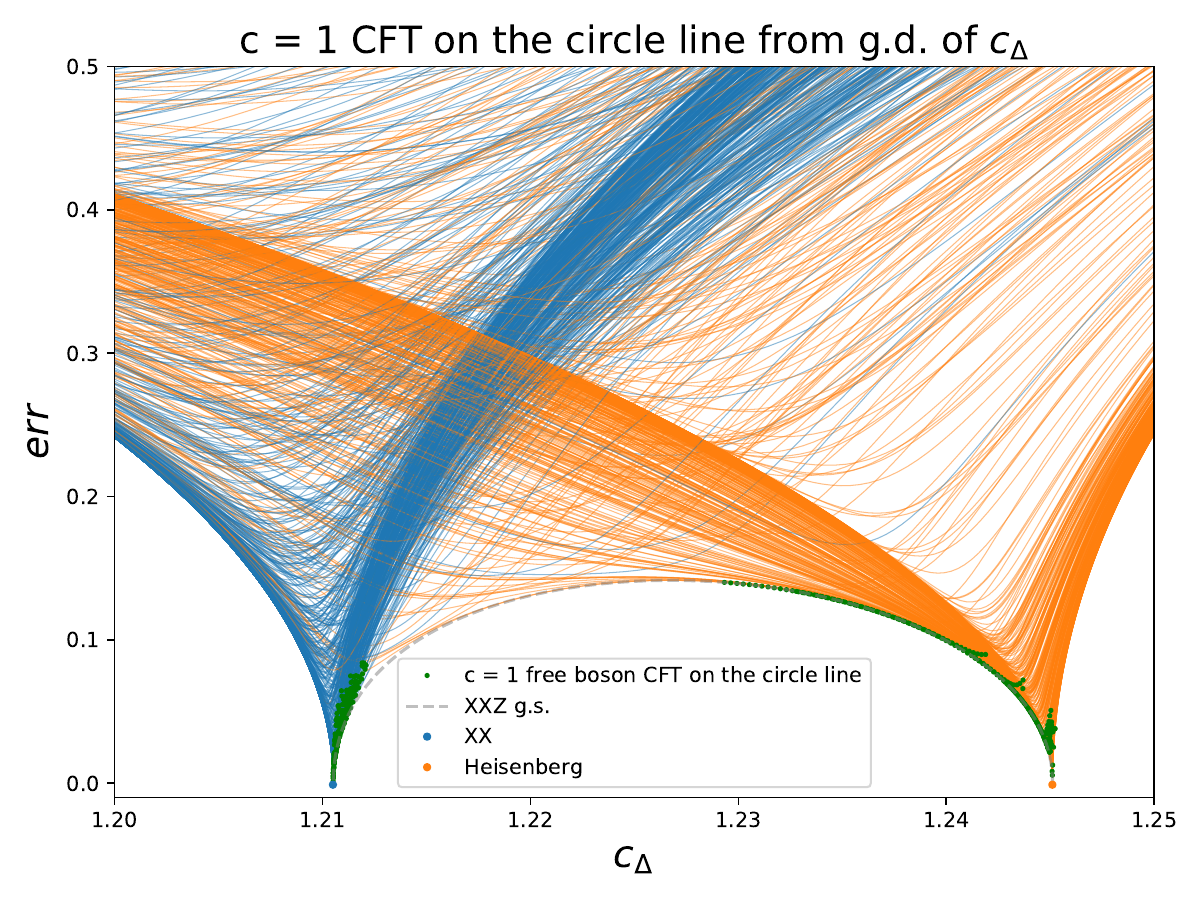}
\includegraphics[width=0.44\linewidth]{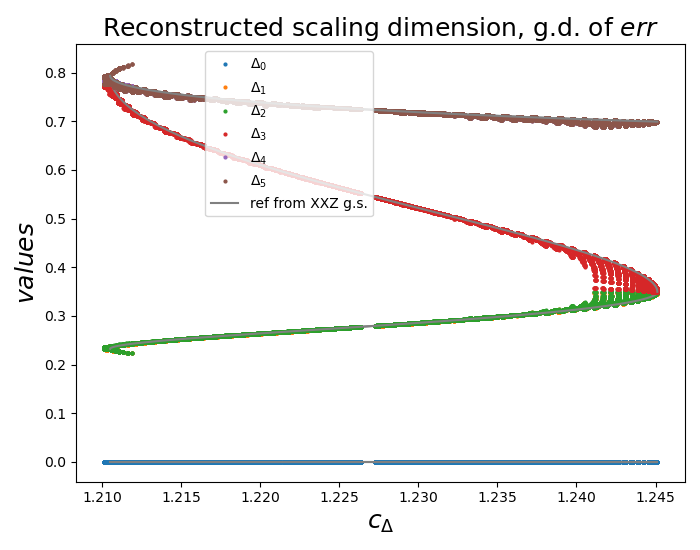}
\includegraphics[width=0.44\linewidth]{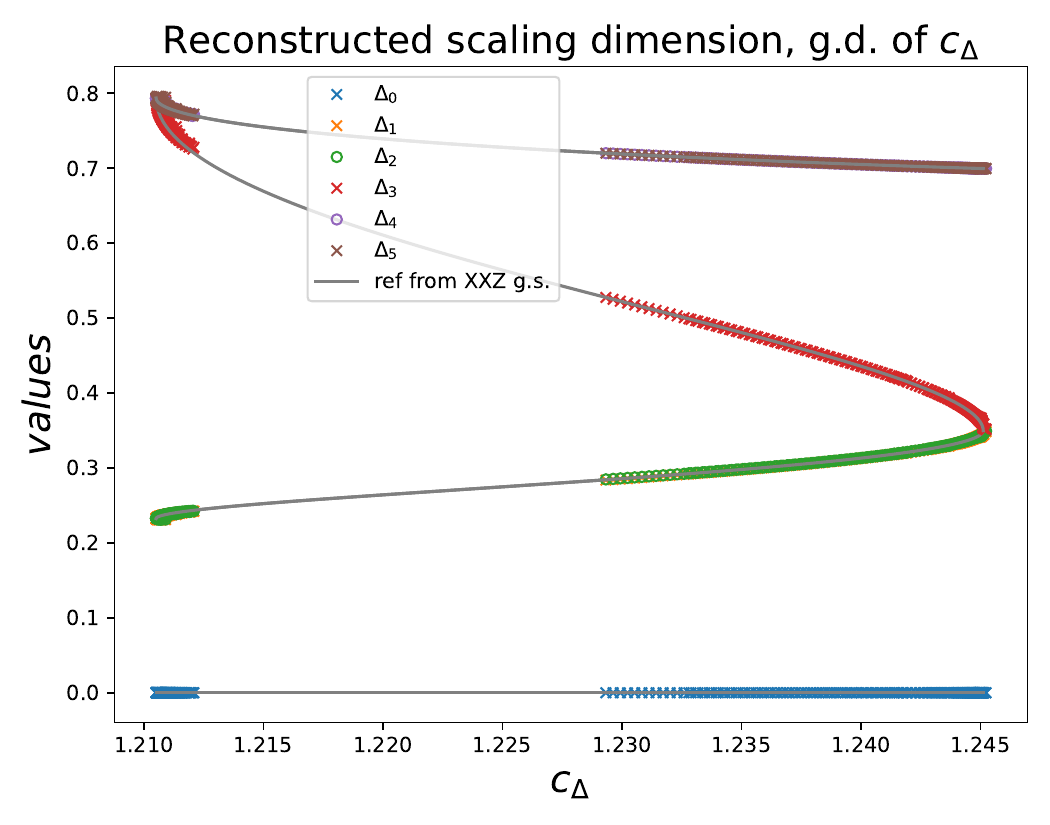}
% \parfig{.23}{figs/c-1-line-gd-c.pdf}
% \parfig{.23}{figs/c-1-line-gd-c-spec.pdf}
\caption{\label{fig:c=1} Top: Gradient descent 
near the $c=1$ fixed points on qubits, XX ($c_\Delta = 1.211$) and Heisenberg ($c_\Delta = 1.245$).  
Top left is gradient descent on the error,
and top right is gradient descent on $c_\Delta$.  
The void between these two points is a finite-size effect.  Also shown (in gray) are $(c_\high, \text{err})$ in groundstates of the XXZ model as $\Delta$ varies from $0$ to $1$, which interpolate between XX and Heisenberg; for intermediate $\Delta$ on 4 sites, there is a finite error.
The green dots are obtained by selecting the states whose $(c_\high, \text{err})$ is close (with deviation less than $10^{-3}$) to that of an XXZ groundstate. 
%The green dots are obtained as follows: 
%First, from the groundstate of $H_\text{XXZ}(\Delta)$ at $L=4$ we compute $ c_\high(\Delta)$ and $err(\Delta)$.  
%Given a value of $c_\high = c_0$ we look for a value of $\Delta$ so that $c_\high(\Delta)$ is close to $c_0$.
% 8 \cdot 10^{-4}.  If the error of that state is also close to $err(\Delta)$, and if there is an approximate degeneracy of the first excited state, then we draw a green dot at $(c_0, err(\Delta))$.  
In this way we match a value of $c_\Delta$ with a value of the parameter $\Delta$ in $H_\text{XXZ}$.
Bottom: The spectra of $H_\text{rec}$ from states at the boundary of the void 
of each of the two top plots, respectively. 
These spectra agree with those of the $c=1$ theory on the circle line and that of the XXZ model \eqref{eq:XXZ} at the values of $\Delta$ determined by the green dots.
}
\end{figure}

{\bf Further discussion.}
What determines the order in which CFTs appear in our search, as we increase the local Hilbert space dimension?
There is a preference for small values of $c$, which, since the reconstructed Hamiltonian always acts on precisely two sites, is consistent with the spirit of the
$c$-$d$ conjecture of \cite{Latorre:2024uqi}.
We observe that CFTs with primaries of
small dimension do not fit in small Hilbert spaces:
In \Cref{fig:c=1} we show the error of the VFPE in the XXZ model as a function of $\Delta$, which determines the radius of the $c=1$ boson; when the radius takes extreme values (near $0$ or $\infty$) there are light primaries and the error grows.
%\JM{Is the figure reference correct?}
This explains the slow appearance of the unitary minimal models in our search as we increase $d$.

\begin{figure}
%\parfig{.23}{figs/fig-qubits-voids.png}
%\parfig{.23}{figs/err-vs-c-group-1.pdf}
\parfig{.45}{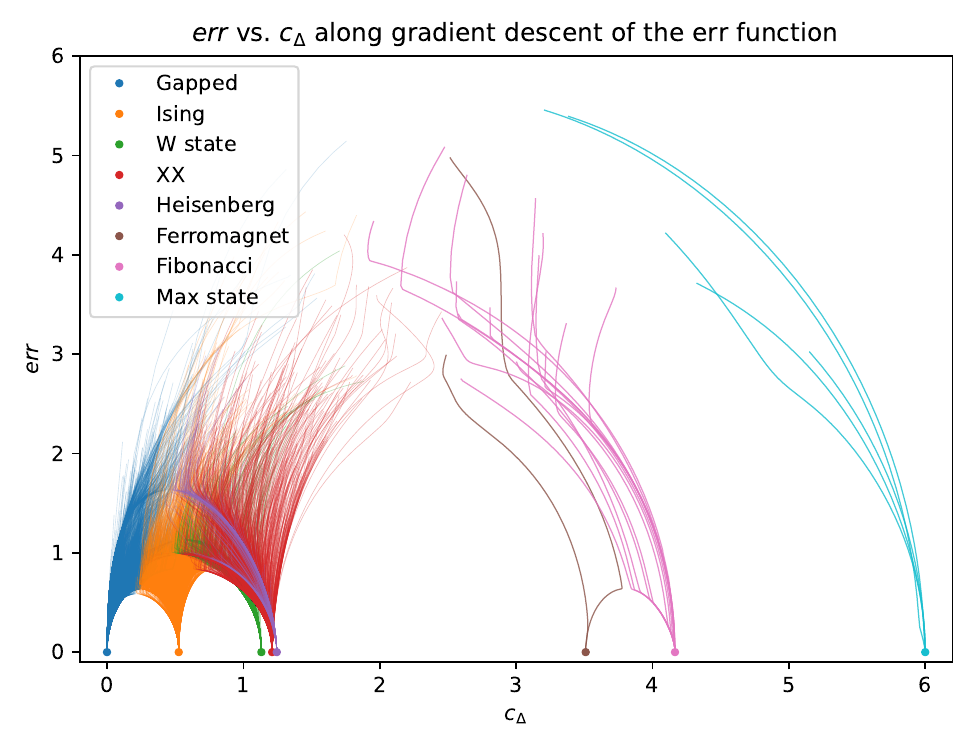}
\parfig{.45}{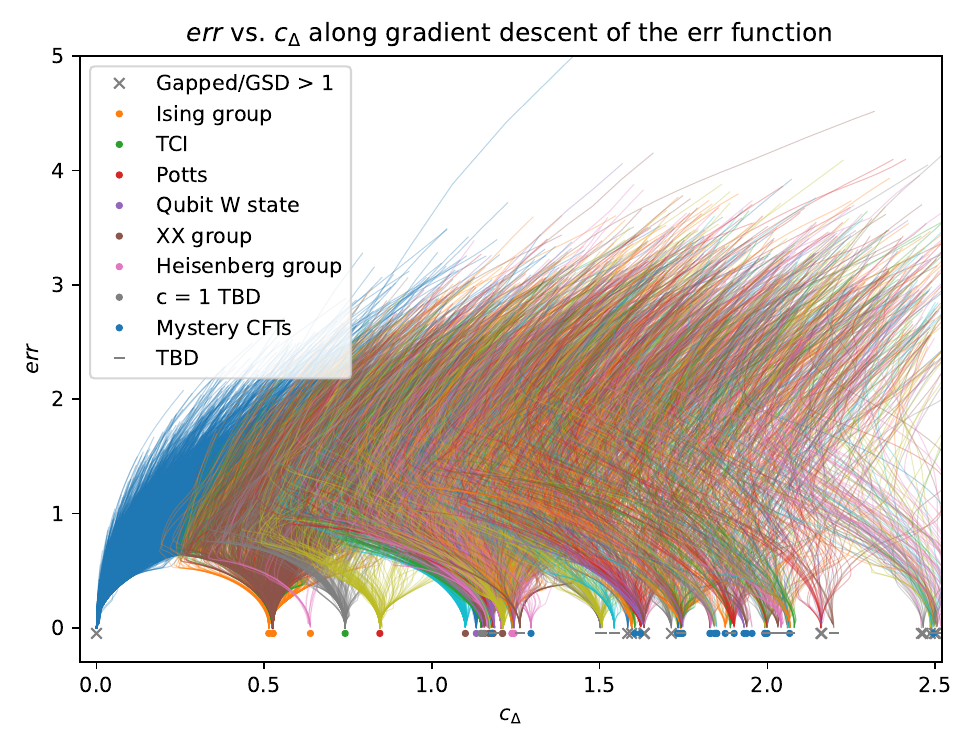}
%\parfig{.23}{figs/err-vs-c-group-1.pdf}
\caption{\label{fig:voids}
Error of the VFPE versus $c_\high$ along the trajectory of the gradient descent of the error function on 4 qubits (top) and 4 qutrits (bottom).
We first do a uniform sampling in the space of translation-invariant and time reversal symmetric states. Then we use the groundstates of their reconstructed Hamiltonians $H_{\text{rec}}$ [Eq.~\eqref{eq:Hrec}] as the initial states in the gradient descent. There are forbidden regions for $c_\high$ when the errors of VFPEs are small.
The names in the figure legends will be explained below.
}
\end{figure}

\begin{figure}
\centering
\parfig{.5}{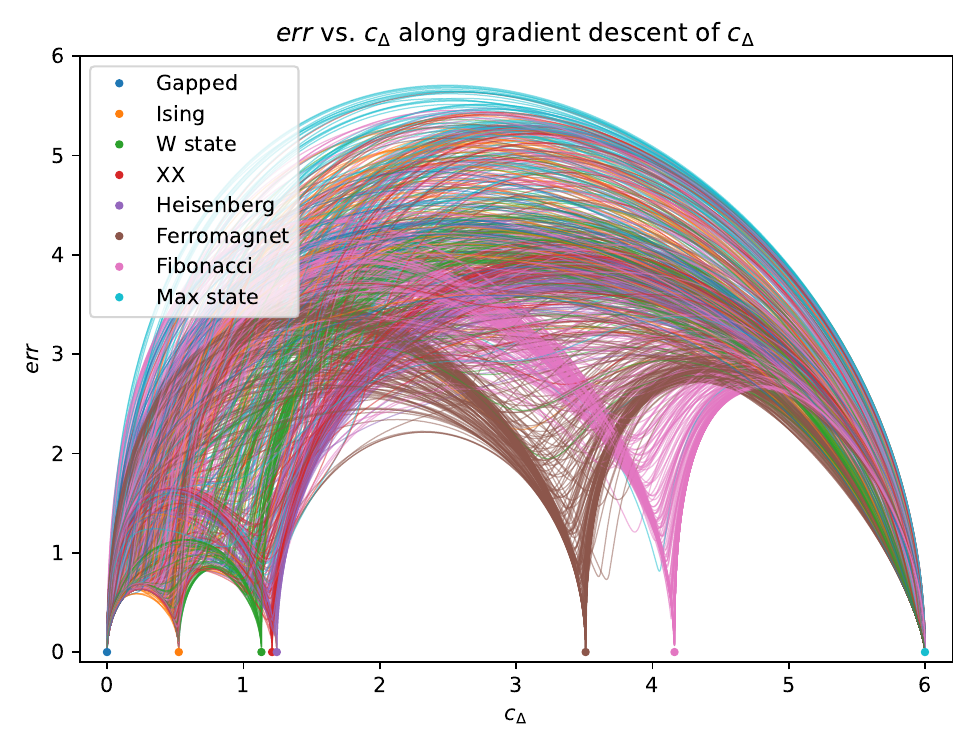}
\caption{
Each curve represents a trajectory of gradient descent on $c_\Delta$, flowing from right to left.
We generate these trajectories by starting with a solution of the VFPE on 4 qubits, adding a small random perturbation, and then perform gradient descent and ascent on $c_\Delta$ to complete each trajectory.  Notice, for example, the flow from XX to Ising.
}
\label{fig:voids2}
\end{figure}

{\bf Approximate solutions of the VFPE.}  In addition to the exact or numerically-exact solutions found above, it is also interesting to study approximate solutions of the VFPE.  One reason we are led to study such approximate solutions is the behavior of CFTs with exactly marginal operators in our search: on four qubits, we find two special points in the $c=1$ moduli space, as isolated critical points of $S_\Delta$, despite the fact that the XXZ model 
\be \label{eq:XXZ}
H_\text{XXZ} = \sum_i \( X_i X_{i+1} + Y_i Y_{i+1}
+ \Delta Z_i Z_{i+1} \) . \ee
is a nearest-neighbor Hamiltonian on qubits that realizes a whole family of $c=1$ CFTs with different radii.
We conclude that the error of the VFPE in these states is a finite-size effect.
By relaxing our search criteria to states with {\it small} error of the VFPE, we can indeed see the whole $c=1$ moduli space (though we have not yet found the Ginsparg archipelago \cite{Ginsparg:1987eb, Ginsparg:1988ui} likely because they have several primaries with small scaling dimensions), see
\Cref{fig:c=1}.
The pattern of approximate solutions also contains information about the space of CFTs.  In particular, one can see voids in the set of states in the plot of error versus $c_\high$ around known forbidden regions, see \Cref{fig:voids,fig:voids2}.
%%% no whitespace allowed in Cref.
There could be some quantum-information-theoretic reason that forbids states in these areas: 
%This potentially suggests a quantum-information-theoretic statement that 
we conjecture that, when $\sigma(\hat{c}_\high) \leq \epsilon $, the values of $c_\high$ are constrained to take a set of specific values with a small window of deviations. 
It is also suggestive that we can observe similar voids in regions for $c>1$ where there is so far no known condition forbidding unitary CFTs. 

\section{Discussion}

{\bf $c_\Delta$ as a $c$-function?}
Starting at a saddle point $\ket{\psi}$ of $c_\Delta$ identified with a CFT, we can parameterize nearby states as
$ \ket{\psi} + \ket{d\psi}$.
In turn we can expand $\ket{d\psi}$ in a basis of eigenstates of $H_\text{rec}^\psi$, which, by the state-operator correspondence of CFT, correspond to operators of definite scaling dimension, $\ket{\Delta_i}$.  Thus, we can ask about the change in $c_\Delta$ upon deforming the state by a given $\ket{\Delta_i}$.
It is natural to ask about a possible relation between the spectrum of $H_\text{rec}^\psi$ and
the spectrum of the Hessian of $c_\Delta$.
The example of CFTs with exactly marginal operators shows that the match is not precise.
However, we observe numerically in examples that the number of relevant operators is equal to the number of downhill directions of $c_\Delta$, that is, negative eigenvalues of its Hessian.
A visceral representation of the structure of this function is given in \Cref{fig:voids2}, where we perturb the solutions of the VFPE and do gradient flow by $c_\high$.  From the plot one can read out flows between fixed points.

{\bf The set of CFTs.}
Does a solution of the VFPE on four sites encode all the data of the associated CFT?
This seems quite believable, though turning this into a universal algorithm remains challenging.
An overly optimistic possibility is that the Hamiltonian reconstructed from such a state, when extended to large system size, would automatically
%have infinite correlation length
realize the CFT in the IR limit.
While this is true for some of the states (such as Heisenberg and XX on qubits), most of the states we find produce $H_\text{rec}$ with a finite correlation length (of order $15-30$ lattice spacing), even when the spectrum of $H_\text{rec}$ for small sizes agrees with the known list of scaling dimensions.  We can systematically improve upon this situation in many ways.  A simple method is: start from the groundstate of $H_\text{rec}^\psi$ on 8 sites of dimension $d$, block the sites into 4 supersites of dimension $d^2$, and use Newton's method to find a lower-error state on the larger local Hilbert space. 

In addition to well-understood states with $c<1$ and $c=1$, in the qutrit search we identify several candidate CFT states
in the range $1 < c < 2$ (see \Cref{tab:qutrits-results-cft-1,tab:qutrits-results-cft-2,tab:qutrits-results-cft-3,tab:qutrits-results-cft-4,tab:qutrits-results-cft-5,tab:qutrits-results-cft-6,tab:qutrits-results-cft-7,tab:qutrits-results-cft-8,tab:qutrits-results-cft-9}).
%\JM{further comments?}
These states clearly deserve further attention.
It would also be interesting to study further the measure on the set of CFTs defined by the canonical measure on the 4-site Hilbert space.
Our search strategy provides a rich source of potential counterexamples to the $c$-$d$ conjecture of \cite{Latorre:2024uqi} --
if we had found a smoking-gun CFT state with $c>2$ in our qutrit search it would have disproved this conjecture.

{\bf Criticality in small spaces.}  We should emphasize an important practical point. $H_\text{rec}$ is a useful tool for learning from small systems:
If you just study the spectrum of an arbitrary lattice Hamiltonian on four sites it doesn't look like CFT, the normalization is arbitrary.  You need to get the stress tensor right to fix the normalization.
But this is generally a very high level in the spectrum and costly to find.
The fact that $H_\text{rec} =  \sum_i \lambda_i K_i$ is completely fixed by $ \sum_i \lambda_i \beta(x) \buildrel{!}\over{=}1 $
means that its normalization is already fixed, as in \eqref{eq:Hrec}.
%On the circle, the correct normalization is
%\be H_\text{rec} = { \cot { \pi \over L } \over L }  \sum_{i=1}^L  \left( K_{[i, i+2]} - K_{[i, i+1]} \right) . \ee
%But $H_{\text{rec}}(\psi)$ has a spectrum much closer to CFT in the sense that it is automatically normalized.
Moreover, $H_\text{rec}$ for the optimized states we find gives the best realization of the continuum CFT in the given Hilbert space, often better than our favorite lattice model.
We have thus provided a method to improve upon a given quantum critical Hamiltonian with the given local Hilbert space dimension.

\begin{figure}
    \centering
\includegraphics[width=.23\textwidth]{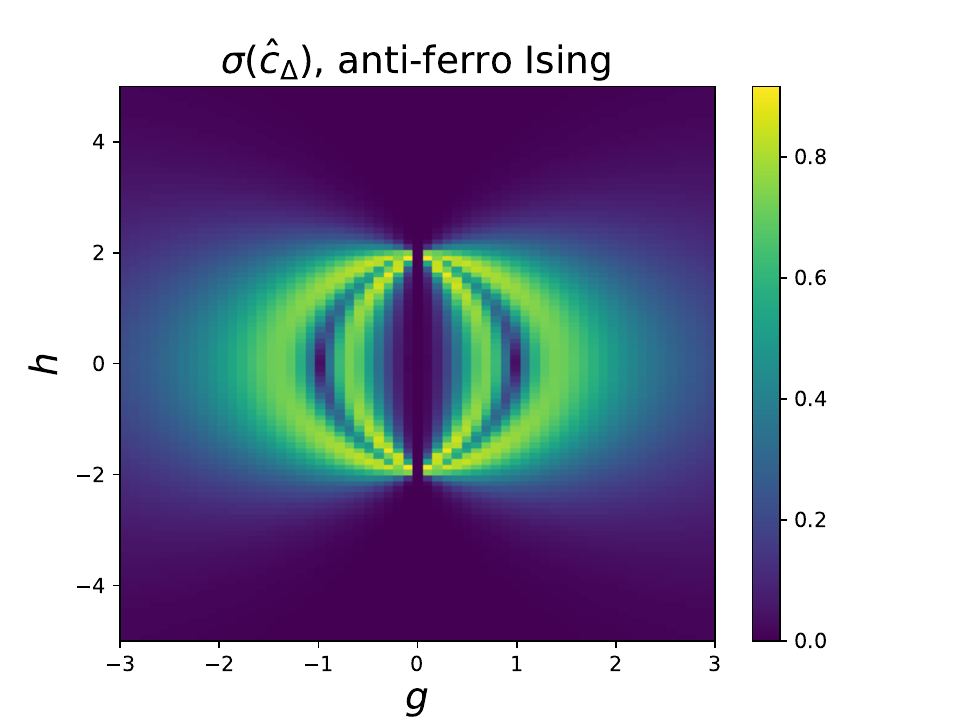}
\includegraphics[width=.23\textwidth]{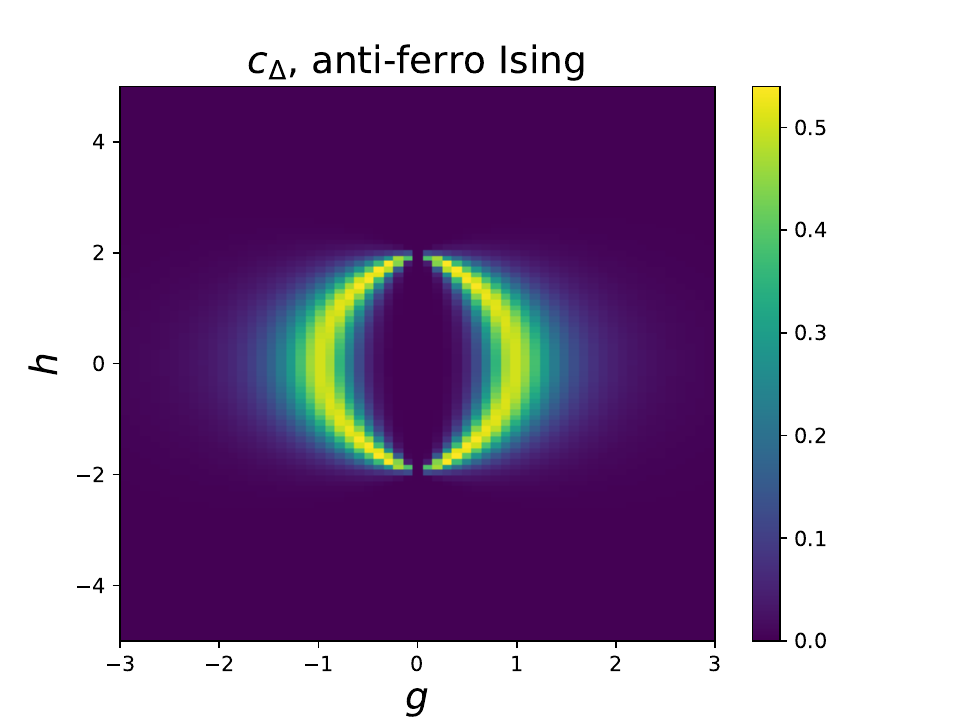}
\caption{The error of the VFPE and the value of $c$ for
the {\it anti}-ferromagnetic Ising chain with transverse and longitudinal fields:
the {\it anti}-ferromagnetic Ising chain with transverse and longitudinal fields:
$
H =  + \sum_{j=1}^L   Z_j Z_{j+1} - h_X \sum_j X_j - h_Z \sum_j Z_j
$
with periodic boundary conditions,
$L = 8$,
and
$ A = \{1,2\}, B = \{3,4\}, C = \{5,6\}$.
We can see two gapped fixed points with $c=0$ separated by a wall of Ising transitions with $c\approx 0.5$.
After a sublattice rotation to map to a ferromagnetic Ising model in a staggered field, it has a hidden Ising symmetry acting by $ \prod_j X_j \otimes $ translation by one site.
}
\label{fig:phase-scan}
\end{figure}

The error of the VFPE also provides a simple practical tool for drawing phase diagrams of a given family of Hamiltonians.
Scanning a phase diagram for minimal error of the VFPE detects fixed points of the RG.
In \Cref{fig:phase-scan} we give an example of this method.
More examples of this application of the VFPE can be found in \cite{janet-and-anran}.

%In this end, we address a philosophical point. Different from the traditional method that one reduce finite size error by increasing the system sizes, the spirit in our method is to optimizing the state in a given system size by reducing the error of the condition which logically defines the nature of the state. For the purpose of studying CFT, with the belief that VFPE\footnote{together with some minor condition that rule out the gapped states} logically derives the defining properties of a CFT (such as the Virasoro algebra relations), by reducing the error of VFPE, one can improve the quality of the state as a finite-size approximation of the CFT.  

\begin{acknowledgments}

We thank Paul Fendley, Tarun Grover, Janet Hung, Yannick Meurice, Daniel Parker, Brandon Rayhaun, Shu-Heng Shao, German Sierra, and the participants of the 2025 Conformal Bootstrap School in S\~ao Paulo, especially Marco Meineri, Jo\~ao Penedones, Leonardo Rastelli, Slava Rychkov, Balt van Rees, Pedro Vieira, for useful discussions and comments.
We thank Shing-Tung Yao for providing inspiration to finish the paper in a timely manner.
DMRG calculations were done using the iTensor library \cite{itensor, itensor-r0.3} in Julia.
This work was supported in part by
funds provided by the U.S. Department of Energy
(D.O.E.) under cooperative research agreement
DE-SC0009919,
and by the Simons Collaboration on Ultra-Quantum Matter, which is a grant from the Simons Foundation (652264, JM).
JM received travel reimbursement from the Simons Foundation;
the terms of this arrangement have been reviewed and approved by the University of California, San Diego in accordance with its conflict of interest policies.

\end{acknowledgments}

\appendix
\renewcommand{\theequation}
{\Alph{section}.\arabic{equation}}

\section{Details of search procedure}
\label{app:search-details}
The goal is to obtain (ideally all) the critical points of the function $c_{\Delta} = \langle \hat{c}_{\Delta} \rangle$ on the space of translation-invariant and time-reversal symmetric states on 4 qudit sites. In the scope of this paper, we choose each site to be a qubit or a qutrit. After we obtain all the critical points, we then want to single out those that correspond to a CFT, which will be referred to as states of interest.

\begin{figure}
  \centering
  \begin{tikzpicture}[every node/.style={inner sep=0,outer sep=0}]
    % Left image
    \node[anchor=north west] (A) at (0,0)
      {\includegraphics[scale=0.44]{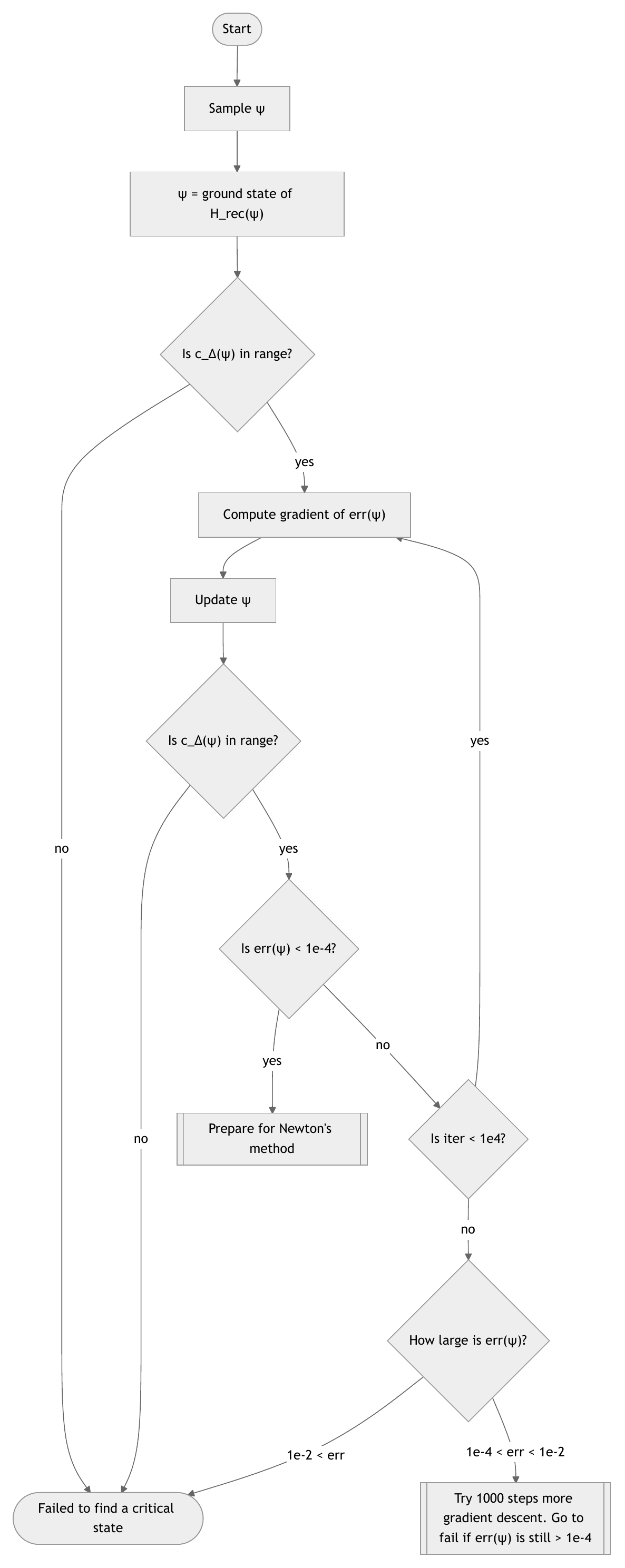}};
    % Place B to the right of A, aligning top edges
    \node[anchor=north west] (B) at (7.8, 0)
      {\includegraphics[scale=0.45]{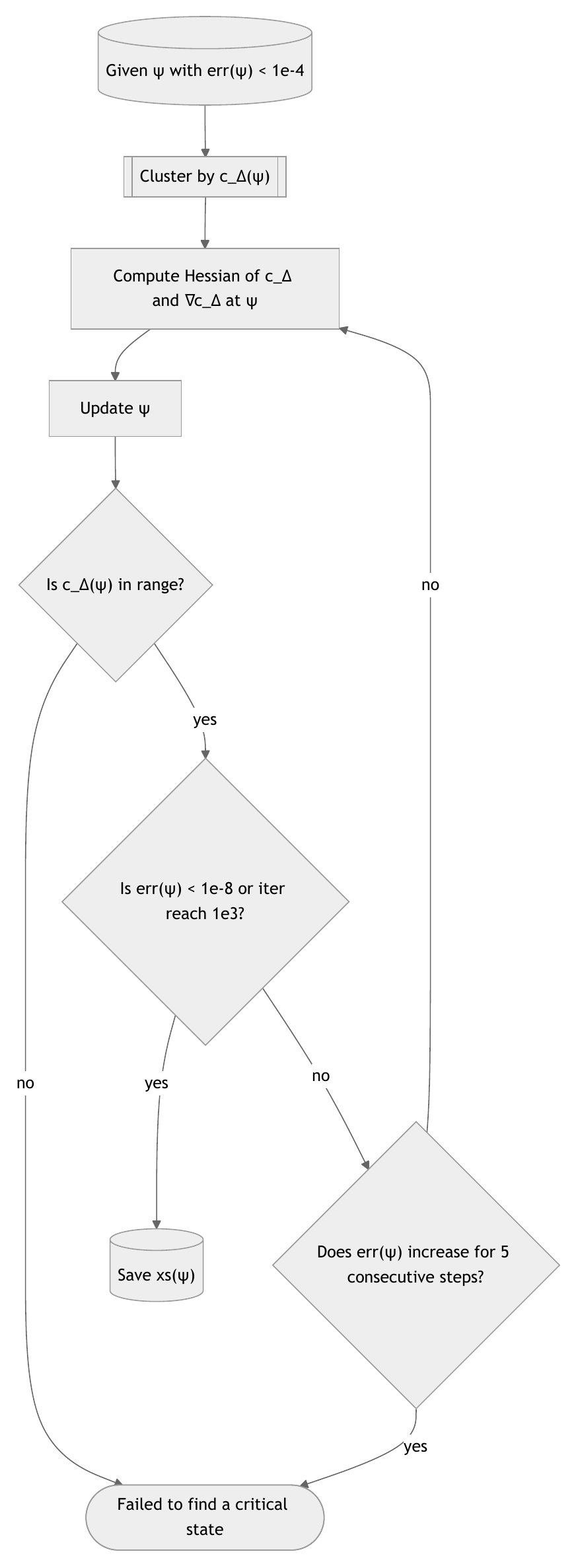}};
    % Place C to the right of B, aligning top edges
    \node[anchor=north west] (C) at (13.1,-16.2)
      {\includegraphics[scale=0.45]{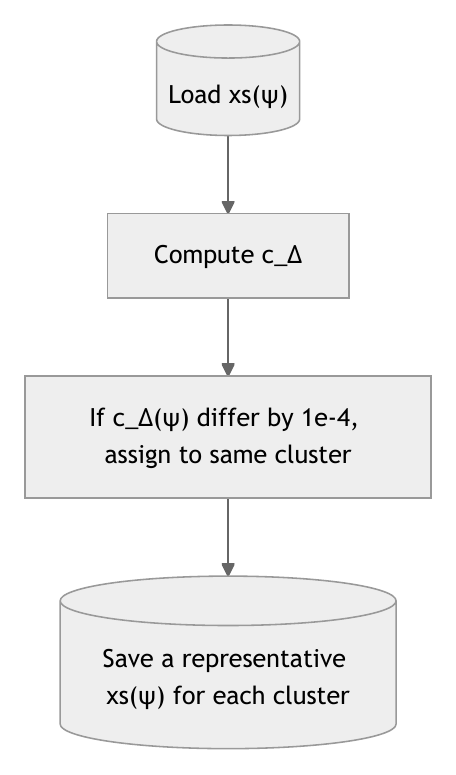}};

    \node[anchor=south] at (3.3,0.2) {Generator};
    \node[anchor=south] at (10.15,0.2) {Newton's method};
    \node[anchor=south] at (14.8,-16) {Clusterer};
  \end{tikzpicture}
  \caption{Flow diagrams of the generator and clusterer.
          The resulting output consists of a list of critical points.}
  \label{fig:flow-chart-find-critical-points}
\end{figure}
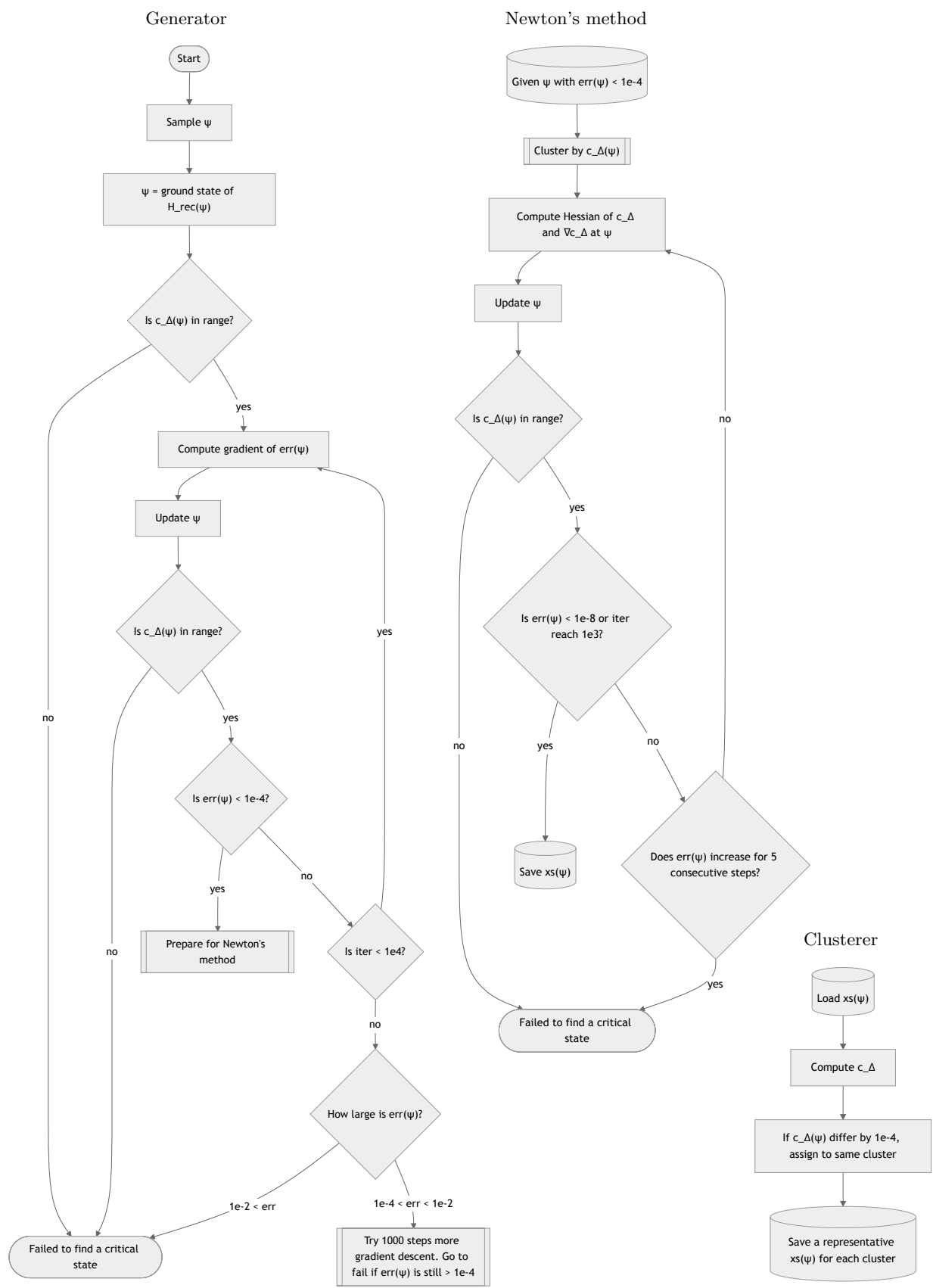

\begin{figure}
  \centering
  \begin{tikzpicture}[every node/.style={inner sep=0,outer sep=0}]
    \node[anchor=north west] (D) at (0,0)
      {\includegraphics[scale=0.46]{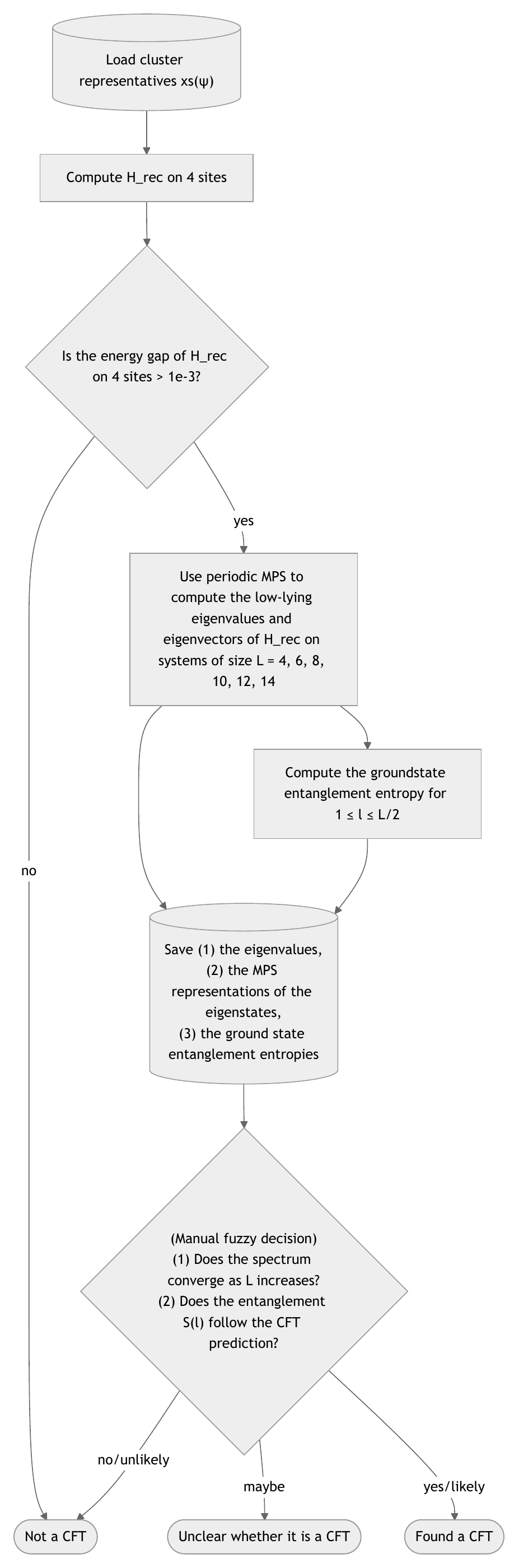}};
    \node[anchor=south] at (2,0.2) {Analyzer};
  \end{tikzpicture}
  \caption{Flow diagram of the analyzer.
          The resulting output consists of a list of CFT candidates
          together with their associated properties,
          including the spectrum.
          The three final outcomes correspond to the three sets of tables at the end of the paper.}
  \label{fig:flow-chart-analyzer}
\end{figure}

The search process contains mainly three components:
(1) generator,
(2) clusterer,
(3) analyzer.
The flow charts can be found in Figures \ref{fig:flow-chart-find-critical-points} and \ref{fig:flow-chart-analyzer}.

The generator is responsible for producing critical states with small error to be analyzed.
This involves sampling from a prior distribution and using gradient descent and Newton's method to find states with small error.

The clusterer groups states with the same physical properties together, allowing for a more efficient analysis by reducing the number of states that need to be considered.

The analyzer performs a detailed examination of the critical states to determine whether they correspond to CFTs. This is achieved through periodic matrix product state (MPS) simulations.
In addition, the procedure enables the estimation of key quantities, including central charges and scaling dimensions.
The states can also be used to estimate OPE data \cite{Zou:2021gja}, though this will be left for future work.

{\bf Generator.}
The generator mainly contains
(1) a sampler that produces initial translation invariant and time-reversal symmetric states.
(2) a gradient descent optimizer that refines the states to minimize the error until the error is less than $10^{-4}$.
(3) a Newton's method optimizer that further refines the states to minimize the error until machine precision.

In between, we employ various filters to throw away states that are not desirable.
This includes states with $c_\Delta \approx 0$ (trivial states) and states that converge too slowly.

(1) We first produce a basis $\{ \ket{s} \}$ for the translation invariant states on 4 sites. A state can be expressed as $\ket{\psi} = \sum_s x_s \ket{s}$, with $x_s$ being real so that the state is time-reversal symmetric.
We then sample the coefficients $x_s$ from a normal distribution and normalize $x_s$, which amounts to uniformly sampling from a hypersphere.

(2) We then perform gradient descent to minimize the error function. Before doing gradient descent directly on the error function $\sigma(\hcd)$, on the initial states $\ket{\psi}_{\text{init}}$ we first construct its reconstructed Hamiltonian and then compute its groundstate $\ket{\psi_0}$. This step can significantly reduce $\sigma(\hcd)$ as well as reduce $\cd$, which can improve the frequency with which we obtain states of interest. Starting from $\ket{\psi_0}$, we perform gradient descent on the error function $\sigma(\hcd)$. To improve computational efficiency, we use numerical differentiation, namely:
\begin{equation}
    \partial_s \sigma = \frac{\sigma(\ket{\psi + \epsilon}_s) - \sigma(\ket{\psi - \epsilon}_s)}{2\epsilon},
\end{equation}
where $\ket{\psi \pm \epsilon}_s = (\ket{\psi} \pm \epsilon \ket{s})/\mathcal{N}_{\pm}$ is the norm-preserving variation with the translation invariant basis $\ket{s}$ and $\mathcal{N}_{\pm}$ is the normalization factor. In practice, we choose $\epsilon = 10^{-5}$. We've compared the numerical differentiation with the analytic differentiation, and we found that choosing $\epsilon$ of the magnitude $10^{-5}$ has the smallest discrepancy. The numerical differentiation is much faster compared with the analytic differentiation. Moreover, since the goal of this step is just to reduce the error to a small amount rather than use it to obtain the exact solution, the numerical error introduced by the numerical differentiation is acceptable. Based on these considerations, we choose to use the numerical differentiation.

With the the gradient of $\sigma(\ket{\psi}) = \sigma[\hcd(\ket{\psi})]$, which is of the form $\nabla \sigma(\ket{\psi}) = \sum_s \partial_s \sigma(\ket{\psi}) \ket{s}$, we can update the state
\begin{equation}
    \ket{\psi}_{\text{new}} = \ket{\psi} - \gamma \cdot \nabla \sigma(\ket{\psi}).
\end{equation}
The step size $\gamma$ is chosen through backtracking line search algorithm. Briefly speaking, we first choose a (relatively large) step size $\gamma_{\text{test}}$ to update the state, examine the decrease of the error.  Then we gradually reduce the step size by a factor $\alpha$: $\gamma_{\text{test}} \to \alpha \gamma_{\text{test}}$, comparing the magnitude of the error decrease with the previous, until we find the best step size in the sequence of tests. From the tests, we found that $\alpha = 0.5$ gave the best performance. Since the landscape we explore is quite rugged, we need to be careful about the step size to ensure convergence. This algorithm will choose a larger step size when the error is large, which speeds up the computation. When the error is small, the number of tests to find the optimal step size will increase and a small step size will be selected.

(3) Once the error becomes sufficiently small $\sigma(\hcd) < 10^{-4}$, the gradient descent method becomes inefficient and we switch to Newton's method to accelerate convergence.\footnote{Intuitively, the Hessian of $c_\Delta$ is believed to be related to the scaling dimensions of the CFT,
  assuming that $c_\Delta$ is related to the Zamolodchikov $c$-function.
This suggests the condition number,
  the ratio between the largest and smallest eigenvalues of the Hessian,
  is rather poor.
Such ill-conditioning may explain the slowness of the current gradient descent method.

Alternatively, one could employ variants of gradient descent with momentum estimation. Whether these variants perform better is deferred to future study.
}
Traditionally, Newton's method is applied near local minima to achieve quadratic convergence. However, it is also known that Newton's method can be applied near saddle points to locate critical points. We adopt this approach to our search procedure.

The Hessian is computed through analytic computations. The time cost for analytically computing the Hessian is similar to that using numerical differentiation. Moreover, since the task in this stage is to obtain an (almost) exact solution up to numerical precision, analytic computation has a relatively small numerical error. If required, we can further reduce the numerical error by changing the machine precision, but we did not find that improving the precision of the numerical estimate of the critical points of the VFPE improved the results for physical quantities.
Once the Hessian $A$ is computed, we can update the state by
\begin{equation}
    \ket{\psi}_{\text{new}} = \ket{\psi} - \beta \cdot A^{-1} \cdot( \nabla \cd),
\end{equation}
where the gradient of $\cd$ can be computed as $\nabla \cd = \hat{c}_{\Delta}\ket{\psi} - c_{\Delta} \ket{\psi}$. Via several tests, we found that $\beta \in (10^{-1}, 10^{-2})$ is the best choice for efficiency. One potential issue is that $A$ might contain zero eigenvalues.
We regulated the computation of the inverse
by adding an identity matrix with a small factor: $A \to A + \delta I$, with $\delta \sim 10^{-7}$. At the end of this step, we typically obtain $\sigma(\hcd) < 10^{-8}$, which is approaching the machine precision for computing $\sigma(\hcd)$.

In practice, Newton's method is time-consuming. To address this, we clustered the results from the gradient descent step and applied Newton's method only to a representative state from each cluster. The clustering algorithm is described below.

{\bf Clusterer.}
The clusterer uses $c_\Delta$ as the key to group states together and outputs one representative state for each group.
When the error $\sigma(\hat{c}_\high)$ is small enough, as shown in Fig.~\ref{fig:voids}, the values of $c_\high$ is clustered. In our numerical results, when $\sigma(\hat{c}_\high) \sim 10^{-4}$, in each cluster the standard deviation of $c_\high$ is of order $10^{-7}$. States in the same cluster are believed to have similar properties, with deviations decreasing as $\sigma(\hat{c}_\high)$ further decreases.

We tested this conjecture by computing the spectrum of $K_{i,i+1}-K_i$ for the states in a cluster and examining their difference.
%The standard deviations are of order [XX].
Since $K_{i,i+1}-K_i$ (the reconstructed Hamiltonian density) are the same up to numerical errors, it is plausible that the reconstructed Hamiltonian Eq.~\eqref{eq:Hrec} will have the same properties.
In fact, in various cases of states of interest, states in the same cluster are simply related to each other by a product of on-site unitaries (up to numerical error), which correspond to the same CFTs.
For example, two states in the Ising cluster on qubits at $c_\high = 0.526$ in \Cref{table:critical-4-qubit-states} with $ \alpha_6 \to - \alpha_6$ are related by the on-site but not translation-invariant unitary $ \psi \to Z_B Z_D \ket{\psi}$.

In extreme settings where $c_\Delta$ attains the maximal allowed value, the states are no longer related by on-site unitaries, nevertheless, the reconstructed Hamiltonian and its groundstate still have the same properties.
We also tested multiple representatives from clusters at larger $c_\high$ and found the same spectrum of the extended $H_\text{rec}$ on $L > 4$ sites.

{\bf Analyzer.}
For each cluster, we pick a representative state and check if its properties are consistent with those of a quantum critical groundstate.
We first employ a simple filter that rejects states whose $H_\text{rec}$ on 4 sites has degenerate ground states.

We then study the low-energy behavior of $H_\text{rec}$ on many sites to approach the thermodynamic limit to extract the universal data.
In particular, we want the energy and the eigenstates.
Due to the complexity in representing quantum states,
we use MPS to represent these low-energy states.
We were able to find the first 8 eigenstates at system sizes $L=4,6,8,10,12,14$ for the first 88 clusters of states, up to $c_\high \approx 2.5$.
For certain states, we also compute $L=16$
  when the spectrum depends on $L \text{mod} 4$.

Additionally, for each system size, we compute the groundstate entanglement entropy of intervals of size $\ell = 1..L/2$.
The computation for each cluster takes about 2.5 hours
on a MacBook Pro 2021 with Apple M1 Max Chip, but is much faster for gapped states.
We also did the computation for a sample of clusters at larger $c_\Delta$.

\section{Details of qutrit search results}
\label{app:search-results-details}

In \Cref{tab:qutrits-results-cft-1,tab:qutrits-results-cft-2,tab:qutrits-results-cft-3,tab:qutrits-results-cft-4,tab:qutrits-results-cft-5,tab:qutrits-results-cft-6,tab:qutrits-results-cft-7,tab:qutrits-results-cft-8,tab:qutrits-results-cft-9,tab:qutrits-results-maybe-cft-1,tab:qutrits-results-maybe-cft-2,tab:qutrits-results-maybe-cft-3,tab:qutrits-results-maybe-cft-4,tab:qutrits-results-no-cft-1,tab:qutrits-results-no-cft-2,tab:qutrits-results-no-cft-3,tab:qutrits-results-no-cft-4}
%In \Cref{tab:qutrits-results-cft-1,tab:qutrits-results-cft-2,tab:qutrits-results-cft-3,tab:qutrits-results-cft-4,tab:qutrits-results-cft-5,tab:qutrits-results-cft-6,tab:qutrits-results-cft-7,tab:qutrits-results-cft-8,tab:qutrits-results-cft-9} and \Cref{tab:qutrits-results-maybe-cft-1,tab:qutrits-results-maybe-cft-2,tab:qutrits-results-maybe-cft-3,tab:qutrits-results-maybe-cft-4} and \Cref{tab:qutrits-results-no-cft-1,tab:qutrits-results-no-cft-2,tab:qutrits-results-no-cft-3,tab:qutrits-results-no-cft-4}
we present an analysis of the first 88 solutions of the VFPE on qutrits that pass the small-system filters (the states themselves can be found in the supplementary material with the arxiv submission).  For each of these states, we study $H_\text{rec}^\psi$ on a chain of length $L \in \{ 4,6,8,10,12,14,(16) \}$ with periodic boundary conditions using the DMRG algorithm (or exact diagonalization for some of the smaller systems).
We find the first 8 eigenstates, and the entanglement entropy $S(\ell)$
in the groundstate of intervals of size
$ \ell = 1.. L/2$.   Based on this information, we group the states into three groups:
states that we believe correspond to CFTs
(\Cref{tab:qutrits-results-cft-1,tab:qutrits-results-cft-2,tab:qutrits-results-cft-3,tab:qutrits-results-cft-4,tab:qutrits-results-cft-5,tab:qutrits-results-cft-6,tab:qutrits-results-cft-7,tab:qutrits-results-cft-8,tab:qutrits-results-cft-9}),
states that we believe do not correspond to CFTs (\Cref{tab:qutrits-results-no-cft-1,tab:qutrits-results-no-cft-2,tab:qutrits-results-no-cft-3,tab:qutrits-results-no-cft-4}),
and states where we cannot draw a conclusion at the moment (\Cref{tab:qutrits-results-maybe-cft-1,tab:qutrits-results-maybe-cft-2,tab:qutrits-results-maybe-cft-3,tab:qutrits-results-maybe-cft-4}).
%The results for each class of states can be found, respectively, in the
%tables \Cref{tab:qutrits-results-cft},
%\Cref{tab:qutrits-results-maybe-cft},
%\Cref{tab:qutrits-results-no-cft}.
For the states that we believe correspond to CFTs, we show
the spectrum of 
the reconstructed dilatation operator, 
${ L \over 2 \pi } H_\text{rec} $, which
(see \eqref{eq:Hrec-and-dimensions}) is normalized so that the numbers should correspond to the list of scaling dimensions of the CFT.
In the middle of the plot, we show the spectrum normalized so that $E_i = 1$, where $i = 1, 2$ or $3$ is chosen so that the spectra collapse best with system size.
On the right, we show the entanglement entropy $S(\ell)$ versus $ \log \sin {\pi \ell \over L} $, whose slope should give $ c/3$ for a CFT state.
In an inset of the plot, we show the values of
\be c_{21} \equiv  3 \( S(\ell=2) - S(\ell=1)\) / (\log\({\sin(2\pi/L) \over \sin(\pi/L) }  \) , \ee
which gives a UV-insensitive estimate of $c$ that is less sensitive to the effects of a potential relevant perturbation than the fits to the slope of $S(\ell)$ up to half the system size.
From the spread of these numbers we can infer the uncertainty in our estimate of the actual central charge of the ideal CFT.

Among the states that we believe are CFTs, we are able to identify
several realizations of the Ising model,
the tricritical Ising model (TCI), the Potts model,
and many realization of CFTs with $c=1$.
Based on the spectra, most of the latter are either the XX point
(where the first two excited states are degenerate at $\Delta = 0.25$)
or the Heisenberg point
(where the first three excited states are degenerate at $\Delta = 0.5$),
though some realize radii in between these two values.
We do not see a clear realization of the orbifold line of $c=1$ theories in exact solutions of the VFPE on (qubits or) qutrits, though representatives will appear for $d=4$ or if we allow finite error.

\begin{figure}
    \centering
    \includegraphics[width=0.9\linewidth]{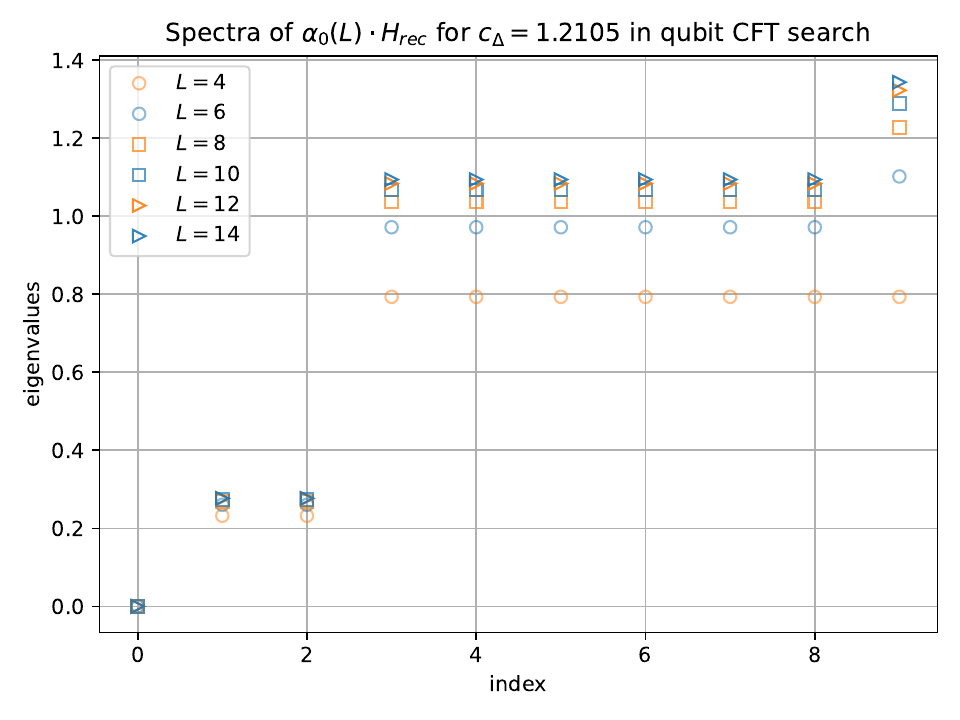}
~~    \includegraphics[width=0.9\linewidth]{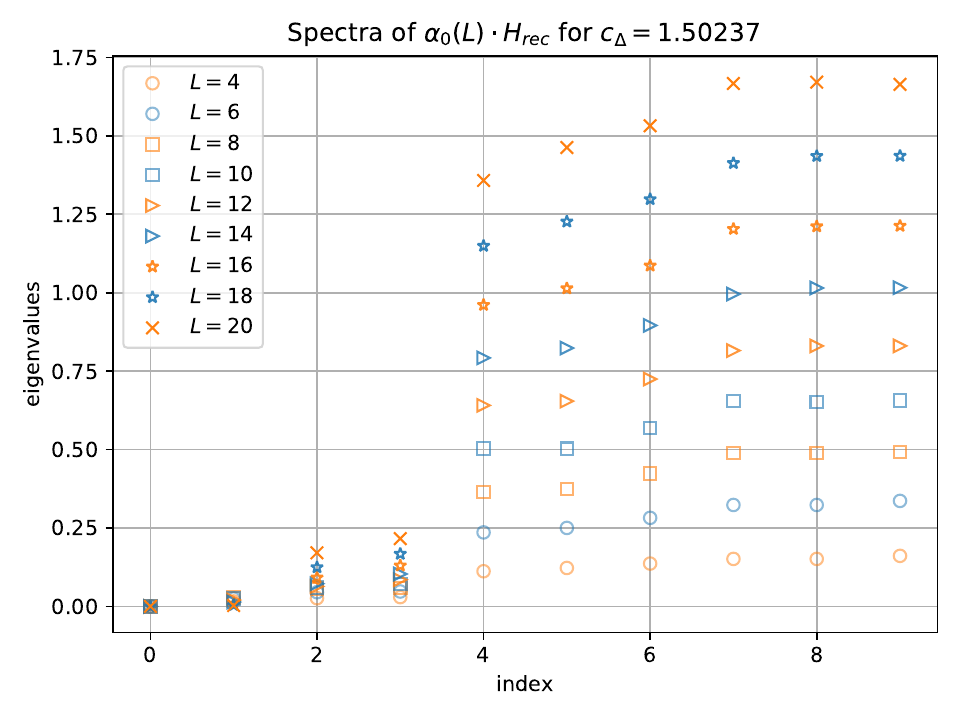}
    \caption{Left: The spectrum of $H_\text{rec} \cdot { L \over 2 \pi}$ from the solution of the VFPE on 4 qubits associated with the XX model.
    The reconstructed Hamiltonian is that of the critical lattice model with infinite correlation length, and the spectrum is converging with system size.
    Right: The spectrum of $H_\text{rec} \cdot { L \over 2 \pi}$ from the solution of the VFPE on 4 qutrits at
    $c_\high = 1.502373$ (labelled `Gapped/GSD$>$1' in \Cref{fig:critical-4-qutrit-states}).
    The spectrum (above the first four states) drifts up linearly with system size.  States with this behavior are excluded from our list of CFT states \Cref{tab:qutrits-results-cft-1,tab:qutrits-results-cft-2,tab:qutrits-results-cft-3,tab:qutrits-results-cft-4,tab:qutrits-results-cft-5,tab:qutrits-results-cft-6,tab:qutrits-results-cft-7,tab:qutrits-results-cft-8,tab:qutrits-results-cft-9}
 and appear instead in
  \Cref{tab:qutrits-results-maybe-cft-1,tab:qutrits-results-maybe-cft-2,tab:qutrits-results-maybe-cft-3,tab:qutrits-results-maybe-cft-4,tab:qutrits-results-no-cft-1,tab:qutrits-results-no-cft-2,tab:qutrits-results-no-cft-3,tab:qutrits-results-no-cft-4}.
 }
    \label{fig:c=1.5}
\end{figure}

\begin{figure}
    \centering
    \includegraphics[width=0.9\linewidth]{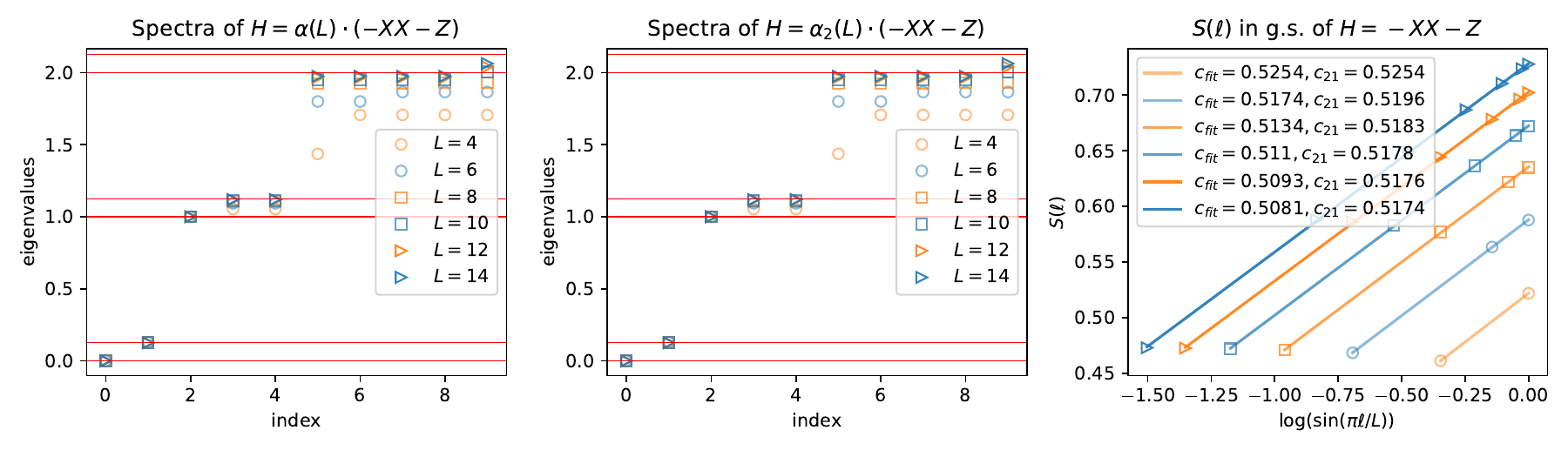}
   \includegraphics[width=0.9\linewidth]{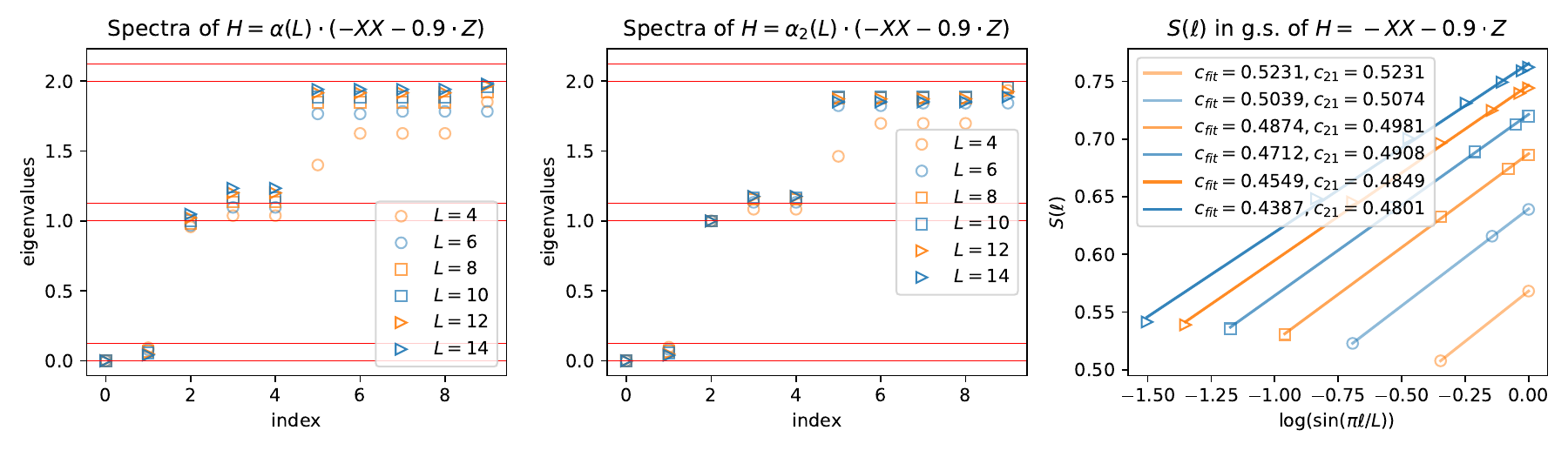}
    \includegraphics[width=0.9\linewidth]{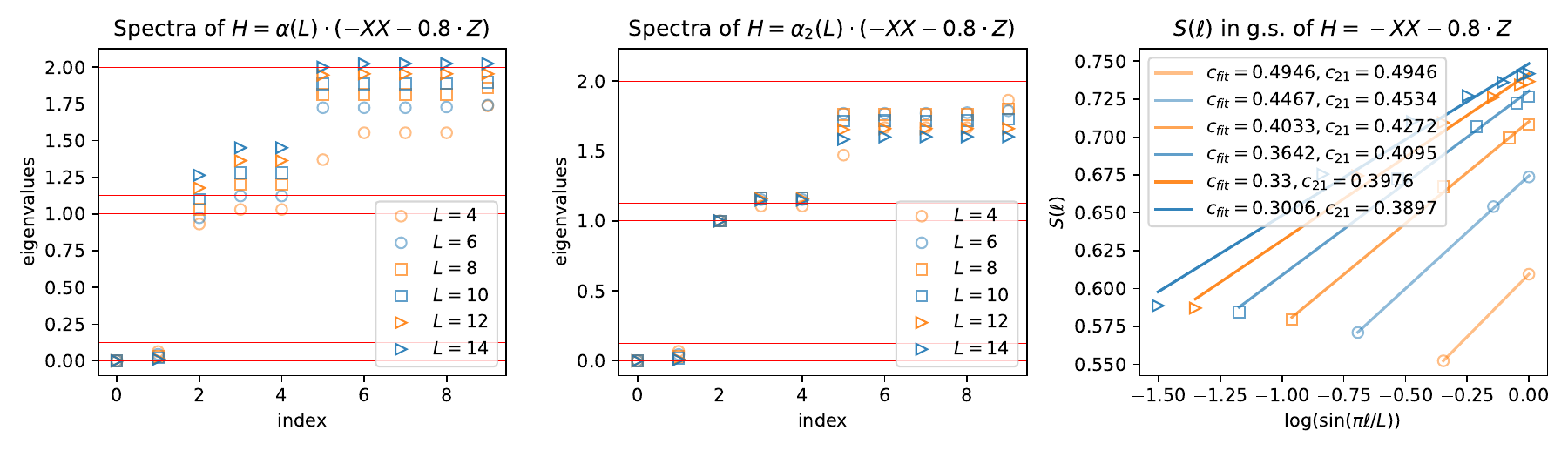}
    \caption{The effects of a relevant perturbation on the low-lying spectrum of a quantum critical lattice model.  Top: the spectrum of the critical transverse-field Ising model $H = - \sum_i X_i X_{i+1} - h \sum_i Z_i $ at $h=1$ at $L=20$ with periodic boundary conditions, from DMRG.
    The normalization is chosen so that the $\epsilon$ state (third eigenstate) has energy $1.0$ at $L=4$.
    Middle: $h = 0.9$.  Bottom: $h=0.8$.
Our purpose in showing this plot is for purposes of comparison with the spectra of $H_\text{rec}$ from states that we identify as CFTs plus a relevant perturbation, such as the ones at $c_\high = 1.50237, 1.50674, 1.74272, 1.74382, 1.74805$, in contrast, for example, to the spectra from the state at $c_\high = 1.60458$ which converges as $L$ grows.
%    \JM{FIXED NORMALIZATION}
        }
    \label{fig:effects-of-relevant-perturbation}
\end{figure}

Farther down the list, we find 25 states that appear to correspond to CFTs with $c>1$, but which we have not yet identified, so we label them `Mystery CFTs'.  Let us describe the criteria by which we selected these states.
There are essentially two: (1) Does the spectrum behave like that of a CFT?
In the leftmost plot of \Cref{fig:effects-of-relevant-perturbation}, we show the finite-size behavior of the spectrum of the critical transverse-field Ising model Hamiltonian $H_\text{TFIM}$.
The normalization is chosen so that at size $L$,
the spectrum is that of $a L H_\text{TFIM} $
where the constant $a$ is chosen so that the state corresponding to $\epsilon$ is at $1$ for $L=4$.
The spectrum is visibly converging to the correct CFT spectrum.
In contrast, when we make a large relevant perturbation of the critical lattice model
(as in the middle and right plots of \Cref{fig:effects-of-relevant-perturbation}),
the spacing between
$E_i(L)$ for fixed level $i$ becomes linear in $L$.  We refer to this as `drift' of the spectrum, and infer that it indicates that the Hamiltonian is gapped.
A direct comparison
a state whose reconstructed spectra 'drift' (and therefore are not included) and one whose spectra do not can be seen in
\Cref{fig:c=1.5}.

(2) The second necessary condition for us to declare a state to be a CFT
is that the entanglement entropy should behave like that of a CFT.  Specifically, the fits of $S(\ell)$ to the expected CFT behavior should have high quality.
For some of our states, the quality of these fits deteriorates at larger system sizes, which we take as a sign that the correlation length is in fact finite, and that $H_\text{rec}$ is the critical Hamiltonian perturbed by a small relevant perturbation.
This interpretation is corroborated by
the fact that when we do infinite DMRG using the VUMPS method, the entanglement entropy of half the system plateaus as a function of bond dimension.  
%Further evidence for this interpretation can be seen in \Cref{fig:open-DMRG}, where
%for one of these states
%we do finite DMRG with open boundary conditions at much larger $L$, and can see the entropy curve leveling off into an area law behavior.

We emphasize that the input to our search has no connection to integrability or to rational CFT.
So, although the CFTs we have managed to identify in our search all happen to be rational for some chiral algebra, we see no reason to expect that this will be the case for all the CFTs we have found.
We have compared the reconstructed spectra with
the list of known (rational) CFTs with $c \in (1,2)$
(as helpfully listed in \cite{Antunes:2025huk}),
but did not yet find a good match.
We note with some disappointment that the reconstructed spectrum of the state
labelled Mystery CFT 1
with $c_\high = 1.29544$,
whose $c$ inferred from entanglement can be estimated as $ 1.15$,
does not match the spectrum found for the candidate irrational CFT of  \cite{Antunes:2025huk}
with the same approximate $c$.

As was the case with the realizations of the Ising and $c=1$ theories, not all of the solutions of the VFPE associated with CFTs appear to correspond to {\it distinct} CFTs.
Thus, some of our Mystery CFTs are assigned the same number.

Both in the examples where we know the correct answer for $c$,
and from our extractions of $c$ from the entanglement, indeed we find that $c_\high$ always overestimates the correct value.
We note that while some of the qutrit states identified as CFTs have $c_\high > 2$, the inferred values of $c$ from the entanglement are all less than $2$, consistent with the $c$-$d$ conjecture.

Amongst the states that we declare not to be CFTs, we do not distinguish between
CFTs with a large relevant perturbation (where the short-distance behavior resembles CFT), and states which appear to have nothing to do with CFT at any scale (such as the $W$ state).

In \Cref{fig:voids} (bottom) and in \Cref{tab:qutrits-results-cft-1}, one can see a solution of the VFPE on qutrits that gives $c_\high \approx 0.63$, a place where no unitary CFT belongs.  Surprisingly, the error can be made quite small, and the result of DMRG shows a state with a large correlation length!  Closer inspection of the reconstructed spectrum shows that
at system size $L$,
%=2,3,4,5,6,7...$,
$H_\text{rec}$ has $\ell_L$
%= 3, 4, 7, 11, 18, 29...$
low-lying states below a large gap, where $\ell_L$ are again the Lucas numbers.
Moreover, these low-lying states take the form of the lowest-energy states in the Ising spectrum.
We infer that the reconstructed Hamiltonian is approximately $ H_\text{rec} \approx \Gamma H_\text{Fib} + H_\text{Ising}$
where $H_\text{Fib}$ is some realization of the Golden chain Hamiltonian \eqref{eq:H_fib} on qutrits, and $\Gamma$ is a large number.  As $L$ grows, the spectrum rapidly becomes identical to the Ising CFT.  Why $c_\high$ on 4 sites for this state is so far off we do not know.

We also studied a sampling of the many states with larger $c_\Delta$. Among them, we did not find any states that look like good candidates for CFTs, consistent with the $c$-$d$ conjecture.

In the tables of results, we define various normalizations of the Hamiltonian,
$H = \alpha_i(L) H_\text{rec}$, with $\alpha_i(L)$ defined as follows: 
$\alpha_0(L) \equiv {L\over 2\pi}$.  
$\alpha_0(L)$ is the predicted theoretical prefactor, such that $D_\text{rec} \equiv \alpha_0(L) H_\text{rec}$ is the reconstructed dilatation operator, whose spectrum is the list of scaling dimensions of the CFT.  
$\alpha_i(L) $ is a prefactor such that the $i+1$-th level of 
the reconstructed Hamiltonian on $L$ sites, 
$\alpha_i(L) H_\text{rec}^L$, is equal to the $i+1$-th level of 
$D_\text{rec}$ on 4 sites.  
The other normalizations are an attempt to see if the spectra for different $L$ collapse.  In the table we show $\alpha_0$ and the best collapse for $i =1..3$.  

\begin{comment}
\begin{figure}
%\parfig{.45}{figs/}
\caption{\label{fig:open-DMRG} Entanglement entropy of sites $1..\ell$ in the groundstate of the reconstructed Hamiltonian extended to $L= ... $ sites, versus $\log \ell$, for various critical qutrit states.
States are constructed by finite DMRG.
We show this plot to explain how we know that our $H_\text{rec}$ constructed from the 4-site state generally is a relevant perturbation of a fixed point, producing a finite correlation length.
}
\end{figure}
\end{comment}

\clearpage

\bibliographystyle{ucsd}
\bibliography{collection}

\begingroup\raggedright\begin{thebibliography}{10}

\bibitem{Belavin:1984vu}
A.~A. Belavin, A.~M. Polyakov, and A.~B. Zamolodchikov, ``{Infinite Conformal Symmetry in Two-Dimensional Quantum Field Theory},'' {\em Nucl. Phys. B} {\bf 241} (1984) 333--380.

\bibitem{Friedan:1983xq}
D.~Friedan, Z.-A. Qiu, and S.~H. Shenker, ``{Conformal Invariance, Unitarity and Two-Dimensional Critical Exponents},'' {\em Phys. Rev. Lett.} {\bf 52} (1984) 1575--1578.

\bibitem{DiFrancesco:1997nk}
P.~Di~Francesco, P.~Mathieu, and D.~Senechal, {\em {Conformal Field Theory}}.
\newblock Graduate Texts in Contemporary Physics. Springer-Verlag, New York, 1997.

\bibitem{Dotsenko:1998gyp}
V.~Dotsenko, J.~L. Jacobsen, M.-A. Lewis, and M.~Picco, ``{Coupled Potts models: Self-duality and fixed point structure},'' {\em Nucl. Phys. B} {\bf 546} (1999) 505--557, \href{http://arxiv.org/abs/cond-mat/9812227}{{\tt cond-mat/9812227}}.

\bibitem{Dotsenko:2001cct}
V.~S. Dotsenko, J.~L. Jacobsen, X.~S. Nguyen, and R.~Santachiara, ``{Universality of coupled Potts models},'' {\em Nucl. Phys. B} {\bf 631} (2002) 426--446, \href{http://arxiv.org/abs/cond-mat/0112120}{{\tt cond-mat/0112120}}.

\bibitem{Antunes:2022vtb}
A.~Antunes and C.~Behan, ``{Coupled Minimal Conformal Field Theory Models Revisited},'' {\em Phys. Rev. Lett.} {\bf 130} (2023), no.~7 071602, \href{http://arxiv.org/abs/2211.16503}{{\tt 2211.16503}}.

\bibitem{Antunes:2024mfb}
A.~Antunes and C.~Behan, ``{Coupled minimal models revisited II: Constraints from permutation symmetry},'' {\em SciPost Phys.} {\bf 18} (2025), no.~4 132, \href{http://arxiv.org/abs/2412.21107}{{\tt 2412.21107}}.

\bibitem{Antunes:2025huk}
A.~Antunes and J.~Rong, ``{Irrational CFTs from coupled anyon chains with non-invertible symmetries?},'' \href{http://arxiv.org/abs/2507.14280}{{\tt 2507.14280}}.

\bibitem{Shi:2018krj}
B.~Shi and Y.-M. Lu, ``{Characterizing topological order by the information convex},'' {\em Phys. Rev. B} {\bf 99} (2019), no.~3 035112, \href{http://arxiv.org/abs/1801.01519}{{\tt 1801.01519}}.

\bibitem{Shi:2018bfb}
B.~Shi, ``{Seeing topological entanglement through the information convex},'' {\em Phys. Rev. Research.} {\bf 1} (2019) 033048, \href{http://arxiv.org/abs/1810.01986}{{\tt 1810.01986}}.

\bibitem{Shi:2019mlt}
B.~Shi, K.~Kato, and I.~H. Kim, ``{Fusion rules from entanglement},'' {\em Annals Phys.} {\bf 418} (2020) 168164, \href{http://arxiv.org/abs/1906.09376}{{\tt 1906.09376}}.

\bibitem{Shi:2019ngw}
B.~Shi, ``{Verlinde formula from entanglement},'' {\em Phys. Rev. Res.} {\bf 2} (2020), no.~2 023132, \href{http://arxiv.org/abs/1911.01470}{{\tt 1911.01470}}.

\bibitem{Shi:2020jxd}
B.~Shi and I.~H. Kim, ``{Domain Wall Topological Entanglement Entropy},'' {\em Phys. Rev. Lett.} {\bf 126} (2021), no.~14 141602, \href{http://arxiv.org/abs/2008.11794}{{\tt 2008.11794}}.

\bibitem{Shi:2020rne}
B.~Shi and I.~H. Kim, ``{Entanglement bootstrap approach for gapped domain walls},'' {\em Phys. Rev. B} {\bf 103} (2021), no.~11 115150, \href{http://arxiv.org/abs/2008.11793}{{\tt 2008.11793}}.

\bibitem{knots-paper}
J.-L. Huang, J.~McGreevy, and B.~Shi, ``{Knots and entanglement},'' \href{http://arxiv.org/abs/2112.08398}{{\tt 2112.08398}}.

\bibitem{paper-one}
B.~Shi, J.-L. Huang, and J.~McGreevy, ``{Remote detectability from entanglement bootstrap I: Kirby's torus trick},'' \href{http://arxiv.org/abs/2301.07119}{{\tt 2301.07119}}.

\bibitem{Lin:2023pvl}
T.-C. Lin and J.~McGreevy, ``{Conformal Field Theory Ground States as Critical Points of an Entropy Function},'' {\em Phys. Rev. Lett.} {\bf 131} (2023), no.~25 251602, \href{http://arxiv.org/abs/2303.05444}{{\tt 2303.05444}}.

\bibitem{Casini:2011kv}
H.~Casini, M.~Huerta, and R.~C. Myers, ``{Towards a derivation of holographic entanglement entropy},'' {\em JHEP} {\bf 05} (2011) 036, \href{http://arxiv.org/abs/1102.0440}{{\tt 1102.0440}}.

\bibitem{Cardy:2016fqc}
J.~Cardy and E.~Tonni, ``{Entanglement hamiltonians in two-dimensional conformal field theory},'' {\em J. Stat. Mech.} {\bf 1612} (2016), no.~12 123103, \href{http://arxiv.org/abs/1608.01283}{{\tt 1608.01283}}.

\bibitem{strength-of-vfpe}
X.~Li, T.-C. Lin, and J.~McGreevy, ``On the strength of the vector fixed-point equation,'' {\em to appear} (2025).

\bibitem{Kim:2024suq}
I.~H. Kim, X.~Li, T.-C. Lin, J.~McGreevy, and B.~Shi, ``{Conformal geometry from entanglement},'' {\em SciPost Physics} {\bf 18} (Mar., 2025) 102, \href{http://arxiv.org/abs/2404.03725}{{\tt 2404.03725}}.

\bibitem{Kim:2024upb}
I.~H. Kim, X.~Li, T.-C. Lin, J.~McGreevy, and B.~Shi, ``{Chiral Virasoro algebra from a single wavefunction},'' {\em Annals Phys.} {\bf 471} (2024) 169849, \href{http://arxiv.org/abs/2403.18410}{{\tt 2403.18410}}.

\bibitem{Zamolodchikov}
A.~B. Zamolodchikov, ``{Irreversibility of the Flux of the Renormalization Group in a 2D Field Theory},'' {\em JETP Lett.} {\bf 43} (1986) 730--732.

\bibitem{Lin:2022jtx}
T.-C. Lin, I.~H. Kim, and M.-H. Hsieh, ``{A new operator extension of strong subadditivity of quantum entropy},'' {\em Lett. Math. Phys.} {\bf 113} (2023), no.~3 68, \href{http://arxiv.org/abs/2211.13372}{{\tt 2211.13372}}.

\bibitem{Ozzello:2025tfu}
Z.~Ozzello and Y.~Meurice, ``{Multipartite entanglement from ditstrings for 1+1D systems},'' \href{http://arxiv.org/abs/2507.14422}{{\tt 2507.14422}}.

\bibitem{zou2020conformal}
Y.~Zou, A.~Milsted, and G.~Vidal, ``Conformal fields and operator product expansion in critical quantum spin chains,'' {\em Physical Review Letters} {\bf 124} (2020), no.~4 040604.

\bibitem{poland2019conformal}
D.~Poland, S.~Rychkov, and A.~Vichi, ``The conformal bootstrap: Theory, numerical techniques, and applications,'' {\em Reviews of Modern Physics} {\bf 91} (2019), no.~1 015002.

\bibitem{Gioia:2023adm}
L.~Gioia and R.~Thorngren, ``{$W$ state is not the unique ground state of any local Hamiltonian},'' \href{http://arxiv.org/abs/2310.10716}{{\tt 2310.10716}}.

\bibitem{Feiguin:2006ydp}
A.~Feiguin, S.~Trebst, A.~W.~W. Ludwig, M.~Troyer, A.~Kitaev, Z.~Wang, and M.~H. Freedman, ``{Interacting anyons in topological quantum liquids: The golden chain},'' {\em Phys. Rev. Lett.} {\bf 98} (2007), no.~16 160409, \href{http://arxiv.org/abs/cond-mat/0612341}{{\tt cond-mat/0612341}}.

\bibitem{Aasen:2016dop}
D.~Aasen, R.~S.~K. Mong, and P.~Fendley, ``{Topological Defects on the Lattice I: The Ising model},'' {\em J. Phys. A} {\bf 49} (2016), no.~35 354001, \href{http://arxiv.org/abs/1601.07185}{{\tt 1601.07185}}.

\bibitem{Aasen:2020jwb}
D.~Aasen, P.~Fendley, and R.~S.~K. Mong, ``{Topological Defects on the Lattice: Dualities and Degeneracies},'' \href{http://arxiv.org/abs/2008.08598}{{\tt 2008.08598}}.

\bibitem{Latorre:2024uqi}
J.~I. Latorre and G.~Sierra, ``{The c\,\ensuremath{-}\,d conjecture},'' {\em J. Stat. Mech.} {\bf 2024} (2024), no.~11 113103, \href{http://arxiv.org/abs/2403.17242}{{\tt 2403.17242}}.

\bibitem{Ginsparg:1987eb}
P.~H. Ginsparg, ``{Curiosities at c = 1},'' {\em Nucl. Phys. B} {\bf 295} (1988) 153--170.

\bibitem{Ginsparg:1988ui}
P.~H. Ginsparg, ``{APPLIED CONFORMAL FIELD THEORY},'' \href{http://arxiv.org/abs/hep-th/9108028}{{\tt hep-th/9108028}}.

\bibitem{janet-and-anran}
A.~Jin and L.-Y. Hung, ``Note on searching for critical lattice models as entropy critical points from strange correlator,'' {\em to appear} (2025).

\bibitem{itensor}
M.~Fishman, S.~R. White, and E.~M. Stoudenmire, ``{The ITensor Software Library for Tensor Network Calculations},'' {\em SciPost Phys. Codebases} (2022) 4, \href{https://scipost.org/10.21468/SciPostPhysCodeb.4}{https://scipost.org/10.21468/SciPostPhysCodeb.4}.

\bibitem{itensor-r0.3}
M.~Fishman, S.~R. White, and E.~M. Stoudenmire, ``{Codebase release 0.3 for ITensor},'' {\em SciPost Phys. Codebases} (2022) 4--r0.3, \href{https://scipost.org/10.21468/SciPostPhysCodeb.4-r0.3}{https://scipost.org/10.21468/SciPostPhysCodeb.4-r0.3}.

\bibitem{Zou:2021gja}
Y.~Zou, ``{Universal information of critical quantum spin chains from wavefunction overlap},'' {\em Phys. Rev. B} {\bf 105} (2022), no.~16 165420, \href{http://arxiv.org/abs/2104.00103}{{\tt 2104.00103}}.

\end{thebibliography}\endgroup

\vfill\eject

\onecolumngrid
\clearpage
\clearpage
\clearpage

\newcommand{\makefilenameyes}[1]{CFT-search-qutrits-dmrg-7-spectra-entropy-plots-0/CFT-search-qutrits-dmrg-spectra-c-#1.pdf}

\newcommand{\rowmacro}[3]{%
  \hline
  #1 & #2 &
  \parbox{.75\textwidth}{%
    \parfig{.75}{\makefilenameyes{#1}} \\
 {\raggedright \footnotesize #3\par}
\vspace{0.5ex}
  }\\
}

\newcommand{\doSomethingIfFour}[1]{%
  \ifnum#1=4
   *
  \fi
}

\doSomethingIfFour{0}

\begin{center}
\begin{table}
\begin{tabular}{|c|c|c|}
  \hline
  $c_\text{high}$ & name & spectra \\ \hline
  \rowmacro{0.51360}{Ising}{  \doSomethingIfFour{0}}
  \rowmacro{0.52445}{$\begin{matrix}\approx \\ \text{qubit Ising} \end{matrix} $}{  \doSomethingIfFour{0}}
  \rowmacro{0.52639}{qubit Ising}{  \doSomethingIfFour{0}}
  \rowmacro{0.63776}{Ising in Fib}{  \doSomethingIfFour{0}}
  \rowmacro{0.74138}{TCI}{  \doSomethingIfFour{0}}
  \rowmacro{0.84475}{Potts}{  \doSomethingIfFour{0}}
  \hline
\end{tabular}
\caption{States on four qutrits that we believe correspond to CFTs (part 1).  
        $\star$ indicates that the spectrum depends on $L$ mod 4.}
\label{tab:qutrits-results-cft-1}
\end{table}
\end{center}

\begin{center}
\begin{table}
\begin{tabular}{|c|c|c|}
  \hline
  $c_\text{high}$ & name & spectra \\ \hline
  \rowmacro{1.09970}{$\begin{matrix} c=1, \\ \text{XX} \end{matrix}$}{  \doSomethingIfFour{4}}
  \rowmacro{1.14620}{$c=1$}{$R \in (R_\text{XX}, R_\text{Heis})$ \doSomethingIfFour{0}}
  \rowmacro{1.14670}{$c=1$}{$R \in (R_\text{XX}, R_\text{Heis})$ \doSomethingIfFour{0}}
  \rowmacro{1.15351}{$c=1$}{Heisenberg with relevant perturbation \doSomethingIfFour{0}}
  \rowmacro{1.15715}{$\begin{matrix} c=1, \\ \text{XX} \end{matrix}$}{  \doSomethingIfFour{0}}
  \rowmacro{1.17090}{$\begin{matrix} c=1, \\ \text{XX} \end{matrix}$}{  \doSomethingIfFour{0}}
  \hline
\end{tabular}
\caption{States on four qutrits that we believe correspond to CFTs (part 2).  
        $\star$ indicates that the spectrum depends on $L$ mod 4.}
\label{tab:qutrits-results-cft-2}
\end{table}
\end{center}

\begin{center}
\begin{table}
\begin{tabular}{|c|c|c|}
  \hline
  $c_\text{high}$ & name & spectra \\ \hline
  \rowmacro{1.17120}{$\begin{matrix} c=1, \\ \text{XX} \end{matrix}$}{$R \in (R_\text{XX}, R_\text{Heis})$ \doSomethingIfFour{0}}
  \rowmacro{1.17384}{$\begin{matrix} c=1, \\ \text{XX} \end{matrix}$}{  \doSomethingIfFour{0}}
  \rowmacro{1.17828}{Mystery CFT 0}{  \doSomethingIfFour{0}}
  \rowmacro{1.17936}{$\begin{matrix} c=1, \\ \text{XX} \end{matrix}$}{  \doSomethingIfFour{0}}
  \rowmacro{1.18098}{$\begin{matrix} c=1, \\ \text{XX} \end{matrix}$}{  \doSomethingIfFour{0}}
  \rowmacro{1.18233}{$\begin{matrix} c=1, \\ \text{XX} \end{matrix}$}{  \doSomethingIfFour{0}}
  \hline
\end{tabular}
\caption{States on four qutrits that we believe correspond to CFTs (part 3).  
        $\star$ indicates that the spectrum depends on $L$ mod 4.}
\label{tab:qutrits-results-cft-3}
\end{table}
\end{center}

\begin{center}
\begin{table}
\begin{tabular}{|c|c|c|}
  \hline
  $c_\text{high}$ & name & spectra \\ \hline
  \rowmacro{1.18272}{$\begin{matrix} c=1, \\ \text{XX} \end{matrix}$}{  \doSomethingIfFour{0}}
  \rowmacro{1.21030}{$\begin{matrix} c=1, \\ \text{XX} \end{matrix}$}{  \doSomethingIfFour{0}}
  \rowmacro{1.21051}{$\begin{matrix} c=1, \\ \text{qubit} \\ \text{XX} \end{matrix}$}{  \doSomethingIfFour{0}}
  \rowmacro{1.23873}{$\begin{matrix} c=1, \\ \text{Heisenberg} \end{matrix}$}{  \doSomethingIfFour{0}}
  \rowmacro{1.23906}{$\begin{matrix} c=1, \\ \text{Heisenberg} \end{matrix}$}{  \doSomethingIfFour{0}}
  \rowmacro{1.24511}{$\begin{matrix} c=1, \\ \text{qubit} \\  \text{Heisenberg}\end{matrix}$}{  \doSomethingIfFour{0}}
  \hline
\end{tabular}
\caption{States on four qutrits that we believe correspond to CFTs (part 4).  
        $\star$ indicates that the spectrum depends on $L$ mod 4.}
\label{tab:qutrits-results-cft-4}
\end{table}
\end{center}

\begin{center}
\begin{table}
\begin{tabular}{|c|c|c|}
  \hline
  $c_\text{high}$ & name & spectra \\ \hline
  \rowmacro{1.29544}{Mystery CFT 1}{  \doSomethingIfFour{0}}
  \rowmacro{1.60459}{Mystery CFT 2}{  \doSomethingIfFour{0}}
  \rowmacro{1.62156}{Mystery CFT 3}{  \doSomethingIfFour{0}}
  \rowmacro{1.73220}{Mystery CFT 4}{  \doSomethingIfFour{0}}
  \rowmacro{1.73405}{Mystery CFT 4'}{Related to previous \doSomethingIfFour{0}}
  \rowmacro{1.74111}{Mystery CFT 5}{  \doSomethingIfFour{0}}
  \hline
\end{tabular}
\caption{States on four qutrits that we believe correspond to CFTs (part 5).  
        $\star$ indicates that the spectrum depends on $L$ mod 4.}
\label{tab:qutrits-results-cft-5}
\end{table}
\end{center}

\begin{center}
\begin{table}
\begin{tabular}{|c|c|c|}
  \hline
  $c_\text{high}$ & name & spectra \\ \hline
  \rowmacro{1.74382}{Mystery CFT 6}{Similar to previous, spectrum is better \doSomethingIfFour{0}}
  \rowmacro{1.74805}{Mystery CFT 6'}{Related to previous \doSomethingIfFour{0}}
  \rowmacro{1.82871}{Mystery CFT 7}{  \doSomethingIfFour{0}}
  \rowmacro{1.83075}{Mystery CFT 8}{  \doSomethingIfFour{4}}
  \rowmacro{1.84206}{Mystery CFT 9}{$E_2=E_3, E_4= E_5$, relevant perturbation visible \doSomethingIfFour{0}}
  \rowmacro{1.84526}{Mystery CFT 10}{$E_2=E_3=E_4$ \doSomethingIfFour{0}}
  \hline
\end{tabular}
\caption{States on four qutrits that we believe correspond to CFTs (part 6).  
        $\star$ indicates that the spectrum depends on $L$ mod 4.}
\label{tab:qutrits-results-cft-6}
\end{table}
\end{center}

\begin{center}
\begin{table}
\begin{tabular}{|c|c|c|}
  \hline
  $c_\text{high}$ & name & spectra \\ \hline
  \rowmacro{1.84851}{Mystery CFT 11}{$E_2=E_3=E_4$, relevant perturbation visible \doSomethingIfFour{0}}
  \rowmacro{1.84926}{Mystery CFT 11'}{$E_2=E_3=E_4$, relevant perturbation visible \doSomethingIfFour{0}}
  \rowmacro{1.87447}{Mystery CFT 11''}{Relevant perturbation visible, same as previous two? \doSomethingIfFour{0}}
  \rowmacro{1.90104}{Mystery CFT 12}{  \doSomethingIfFour{4}}
  \rowmacro{1.93044}{Mystery CFT 12'}{Related to previous \doSomethingIfFour{4}}
  \rowmacro{1.93838}{Mystery CFT 13}{  \doSomethingIfFour{4}}
  \hline
\end{tabular}
\caption{States on four qutrits that we believe correspond to CFTs (part 7).  
        $\star$ indicates that the spectrum depends on $L$ mod 4.}
\label{tab:qutrits-results-cft-7}
\end{table}
\end{center}

\begin{center}
\begin{table}
\begin{tabular}{|c|c|c|}
  \hline
  $c_\text{high}$ & name & spectra \\ \hline
  \rowmacro{1.95423}{Mystery CFT 14}{  \doSomethingIfFour{0}}
  \rowmacro{1.99222}{Mystery CFT 15}{$E_2=E_3=E_4=E_5=E_6$ \doSomethingIfFour{0}}
  \rowmacro{1.99325}{Mystery CFT 16}{  \doSomethingIfFour{4}}
  \rowmacro{1.99972}{Mystery CFT 17}{  \doSomethingIfFour{4}}
  \rowmacro{2.06646}{Mystery CFT 18}{  \doSomethingIfFour{4}}
  \rowmacro{2.08015}{Mystery CFT 19}{  \doSomethingIfFour{4}}
  \hline
\end{tabular}
\caption{States on four qutrits that we believe correspond to CFTs (part 8).  
        $\star$ indicates that the spectrum depends on $L$ mod 4.}
\label{tab:qutrits-results-cft-8}
\end{table}
\end{center}

\begin{center}
\begin{table}
\begin{tabular}{|c|c|c|}
  \hline
  $c_\text{high}$ & name & spectra \\ \hline
  \rowmacro{2.49527}{Mystery CFT 20}{  \doSomethingIfFour{4}}
  \hline
\end{tabular}
\caption{States on four qutrits that we believe correspond to CFTs (part 9).  
        $\star$ indicates that the spectrum depends on $L$ mod 4.}
\label{tab:qutrits-results-cft-9}
\end{table}
\end{center}

\begin{comment}
\begin{center}
\begin{table}
\begin{tabular}{|c|c|c|}
     \hline
  $ c_\high$ & name & spectra of CFT states %& comments
%%%%%%%% BEGIN
\\ \hline
\rowmacro{0.5136}{Ising}{commentary}
\rowmacro{0.52445}{Ising}{commentary}
\rowmacro{0.5136}{Potts}{commentary}
\rowmacro{0.5136}{Potts}{commentary}
\rowmacro{0.5136}{Potts}{commentary}
\rowmacro{0.5136}{Potts}{commentary}
%%%%%%%%%%%%%% END
 \hline
\end{tabular}
\caption{\label{tab:qutrits-results} Filtered results of the search on 4 qutrits.}
\end{table}
\end{center}
\end{comment}

%%%%%%%% FOR MAYBE
\newcommand{\makefilenamemaybe}[1]{CFT-search-qutrits-dmrg-7-spectra-entropy-plots-1/CFT-search-qutrits-dmrg-spectra-c-#1.pdf}

\newcommand{\rowmacroone}[2]{%
  \hline
  #1  &
    \parfig{.45}{\makefilenamemaybe{#1}}
&  \parbox{.25\textwidth}{%
 {\footnotesize #2}
  }
  \\
}
\begin{center}
\begin{table}
\begin{tabular}{|c|c|c|}
  \hline
  $c_\text{high}$ & spectra & comments \\ \hline
  \rowmacroone{1.15139}{ }
  \rowmacroone{1.18670}{GSD = 2}
  \rowmacroone{1.26197}{Perhaps a big relevant perturbation of a CFT.  Larger $L$ has smaller $S$.}
  \rowmacroone{1.50674}{Spectrum is drifting?}
  \rowmacroone{1.54380}{Larger $L$ has smaller $S$}
  \rowmacroone{1.74272}{Maybe the spectrum is converging?}
  \hline
\end{tabular}
\caption{States on four qutrits where we cannot tell whether they correspond to CFTs (part 1).  
        $\star$ indicates that the spectrum depends on $L$ mod 4.}
\label{tab:qutrits-results-maybe-cft-1}
\end{table}
\end{center}

\begin{center}
\begin{table}
\begin{tabular}{|c|c|c|}
  \hline
  $c_\text{high}$ & spectra & comments \\ \hline
  \rowmacroone{1.88979}{ }
  \rowmacroone{1.99557}{$E_2=E_3=E_4=E_5$, $E_6=E_7=E_8$}
  \rowmacroone{1.99790}{Related to previous}
  \rowmacroone{1.99834}{Related to previous}
  \rowmacroone{1.99851}{Related to previous}
  \rowmacroone{2.00000}{ }
  \hline
\end{tabular}
\caption{States on four qutrits where we cannot tell whether they correspond to CFTs (part 2).  
        $\star$ indicates that the spectrum depends on $L$ mod 4.}
\label{tab:qutrits-results-maybe-cft-2}
\end{table}
\end{center}

\begin{center}
\begin{table}
\begin{tabular}{|c|c|c|}
  \hline
  $c_\text{high}$ & spectra & comments \\ \hline
  \rowmacroone{2.03145}{ }
  \rowmacroone{2.05983}{ }
  \rowmacroone{2.06179}{Related to previous}
  \rowmacroone{2.06193}{Related to previous}
  \rowmacroone{2.06802}{Related to previous}
  \rowmacroone{2.19797}{Same as previous}
  \hline
\end{tabular}
\caption{States on four qutrits where we cannot tell whether they correspond to CFTs (part 3).  
        $\star$ indicates that the spectrum depends on $L$ mod 4.}
\label{tab:qutrits-results-maybe-cft-3}
\end{table}
\end{center}

\begin{center}
\begin{table}
\begin{tabular}{|c|c|c|}
  \hline
  $c_\text{high}$ & spectra & comments \\ \hline
  \rowmacroone{2.49499}{ }
  \rowmacroone{2.50322}{ }
  \hline
\end{tabular}
\caption{States on four qutrits where we cannot tell whether they correspond to CFTs (part 4).  
        $\star$ indicates that the spectrum depends on $L$ mod 4.}
\label{tab:qutrits-results-maybe-cft-4}
\end{table}
\end{center}

%%%%%%%% FOR NO
\newcommand{\makefilenameno}[1]{CFT-search-qutrits-dmrg-7-spectra-entropy-plots-2/CFT-search-qutrits-dmrg-spectra-c-#1.pdf}

\newcommand{\rowmacrotwo}[2]{%
  \hline
  #1  &
      \parfig{.45}{\makefilenameno{#1}}
&  \parbox{.25\textwidth}{%
 {\footnotesize #2}
  }\\
}

\begin{center}
\begin{table}
\begin{tabular}{|c|c|c|}
  \hline
  $c_\text{high}$ & spectra & comments \\ \hline
  \rowmacroone{1.13233}{qubit W state}
  \rowmacroone{1.50237}{Spectrum is drifting.  The number of low-lying states (below a gap) is $\ell_L$.}
  \rowmacroone{1.58353}{missing}
  \rowmacroone{1.59520}{Perturbation of a CFT into a gapped phase with GSD}
  \rowmacroone{1.63246}{Gapped}
  \rowmacroone{1.63298}{Gapped, GSD=4 for L=0 mod 4, GSD=2 for L=2 mod 4.}
  \hline
\end{tabular}
\caption{States on four qutrits that we believe do not correspond to CFTs (part 1).  
        $\star$ indicates that the spectrum depends on $L$ mod 4.}
\label{tab:qutrits-results-no-cft-1}
\end{table}
\end{center}

\begin{center}
\begin{table}
\begin{tabular}{|c|c|c|}
  \hline
  $c_\text{high}$ & spectra & comments \\ \hline
  \rowmacroone{1.63357}{Gapped, GSD=4}
  \rowmacroone{1.71387}{Gapped, related to previous}
  \rowmacroone{1.73428}{Spectrum is drifting}
  \rowmacroone{2.04119}{GSD}
  \rowmacroone{2.15965}{GSD}
  \rowmacroone{2.16231}{Same as previous}
  \hline
\end{tabular}
\caption{States on four qutrits that we believe do not correspond to CFTs (part 2).  
        $\star$ indicates that the spectrum depends on $L$ mod 4.}
\label{tab:qutrits-results-no-cft-2}
\end{table}
\end{center}

\begin{center}
\begin{table}
\begin{tabular}{|c|c|c|}
  \hline
  $c_\text{high}$ & spectra & comments \\ \hline
  \rowmacroone{2.45903}{Gapped: spectrum collapses, but drifts away.  Entropy is area law.}
  \rowmacroone{2.46425}{The DMRG routine failed to find the correct eigenstates at L=14.}
  \rowmacroone{2.48588}{GSD = 2}
  \rowmacroone{2.49803}{GSD = 4}
  \rowmacroone{2.49947}{Gapped}
  \rowmacroone{2.50198}{GSD = 2}
  \hline
\end{tabular}
\caption{States on four qutrits that we believe do not correspond to CFTs (part 3).  
        $\star$ indicates that the spectrum depends on $L$ mod 4.}
\label{tab:qutrits-results-no-cft-3}
\end{table}
\end{center}

\begin{center}
\begin{table}
\begin{tabular}{|c|c|c|}
  \hline
  $c_\text{high}$ & spectra & comments \\ \hline
  \rowmacroone{2.51738}{Spectrum is drifting.}
  \hline
\end{tabular}
\caption{States on four qutrits that we believe do not correspond to CFTs (part 4).  
        $\star$ indicates that the spectrum depends on $L$ mod 4.}
\label{tab:qutrits-results-no-cft-4}
\end{table}
\end{center}

\end{document}